\documentclass[aps,prd,reprint,nofootinbib,superscriptaddress, floatfix, amsmath,amssym]{revtex4-2}

\usepackage{graphicx}
\usepackage{dcolumn}
\usepackage{bm}
\usepackage{hyperref}
\usepackage{multirow}

\begin{document}

\title{Measurement of the inelasticity distribution of neutrino-nucleon interactions for $\mathbf{80~GeV<E_{\nu}<560~GeV}$ with IceCube DeepCore}
\affiliation{III. Physikalisches Institut, RWTH Aachen University, D-52056 Aachen, Germany}
\affiliation{Department of Physics, University of Adelaide, Adelaide, 5005, Australia}
\affiliation{Dept. of Physics and Astronomy, University of Alaska Anchorage, 3211 Providence Dr., Anchorage, Alaska 99508, USA}
\affiliation{Dept. of Physics, University of Texas at Arlington, 502 Yates Street, Science Hall Rm 108, Box 19059, Arlington, Texas 76019, USA}
\affiliation{School of Physics and Center for Relativistic Astrophysics, Georgia Institute of Technology, Atlanta, Georgia 30332, USA}
\affiliation{Dept. of Physics, Southern University, Baton Rouge, Louisiana 70813, USA}
\affiliation{Dept. of Physics, University of California, Berkeley, California 94720, USA}
\affiliation{Lawrence Berkeley National Laboratory, Berkeley, California 94720, USA}
\affiliation{Institut f{\"u}r Physik, Humboldt-Universit{\"a}t zu Berlin, D-12489 Berlin, Germany}
\affiliation{Fakult{\"a}t f{\"u}r Physik {\&} Astronomie, Ruhr-Universit{\"a}t Bochum, D-44780 Bochum, Germany}
\affiliation{Universit{\'e} Libre de Bruxelles, Science Faculty CP230, B-1050 Brussels, Belgium}
\affiliation{Vrije Universiteit Brussel (VUB), Dienst ELEM, B-1050 Brussels, Belgium}
\affiliation{Dept. of Physics, Simon Fraser University, Burnaby, British Columbia V5A 1S6, Canada}
\affiliation{Department of Physics and Laboratory for Particle Physics and Cosmology, Harvard University, Cambridge, Massachusetts 02138, USA}
\affiliation{Dept. of Physics, Massachusetts Institute of Technology, Cambridge, Massachusetts 02139, USA}
\affiliation{Dept. of Physics and The International Center for Hadron Astrophysics, Chiba University, Chiba 263-8522, Japan}
\affiliation{Department of Physics, Loyola University Chicago, Chicago, Illinois 60660, USA}
\affiliation{Dept. of Physics and Astronomy, University of Canterbury, Private Bag 4800, Christchurch, New Zealand}
\affiliation{Dept. of Physics, University of Maryland, College Park, Maryland 20742, USA}
\affiliation{Dept. of Astronomy, Ohio State University, Columbus, Ohio 43210, USA}
\affiliation{Dept. of Physics and Center for Cosmology and Astro-Particle Physics, Ohio State University, Columbus, Ohio 43210, USA}
\affiliation{Niels Bohr Institute, University of Copenhagen, DK-2100 Copenhagen, Denmark}
\affiliation{Dept. of Physics, TU Dortmund University, D-44221 Dortmund, Germany}
\affiliation{Dept. of Physics and Astronomy, Michigan State University, East Lansing, Michigan 48824, USA}
\affiliation{Dept. of Physics, University of Alberta, Edmonton, Alberta, T6G 2E1, Canada}
\affiliation{Erlangen Centre for Astroparticle Physics, Friedrich-Alexander-Universit{\"a}t Erlangen-N{\"u}rnberg, D-91058 Erlangen, Germany}
\affiliation{Physik-department, Technische Universit{\"a}t M{\"u}nchen, D-85748 Garching, Germany}
\affiliation{D{\'e}partement de physique nucl{\'e}aire et corpusculaire, Universit{\'e} de Gen{\`e}ve, CH-1211 Gen{\`e}ve, Switzerland}
\affiliation{Dept. of Physics and Astronomy, University of Gent, B-9000 Gent, Belgium}
\affiliation{Dept. of Physics and Astronomy, University of California, Irvine, California 92697, USA}
\affiliation{Karlsruhe Institute of Technology, Institute for Astroparticle Physics, D-76021 Karlsruhe, Germany}
\affiliation{Karlsruhe Institute of Technology, Institute of Experimental Particle Physics, D-76021 Karlsruhe, Germany}
\affiliation{Dept. of Physics, Engineering Physics, and Astronomy, Queen's University, Kingston, Ontario K7L 3N6, Canada}
\affiliation{Department of Physics {\&} Astronomy, University of Nevada, Las Vegas, Nevada 89154, USA}
\affiliation{Nevada Center for Astrophysics, University of Nevada, Las Vegas, Nevada 89154, USA}
\affiliation{Dept. of Physics and Astronomy, University of Kansas, Lawrence, Kansas 66045, USA}
\affiliation{Centre for Cosmology, Particle Physics and Phenomenology - CP3, Universit{\'e} catholique de Louvain, Louvain-la-Neuve, Belgium}
\affiliation{Department of Physics, Mercer University, Macon, Georgia 31207-0001, USA}
\affiliation{Dept. of Astronomy, University of Wisconsin{\textemdash}Madison, Madison, Wisconsin 53706, USA}
\affiliation{Dept. of Physics and Wisconsin IceCube Particle Astrophysics Center, University of Wisconsin{\textemdash}Madison, Madison, Wisconsin 53706, USA}
\affiliation{Institute of Physics, University of Mainz, Staudinger Weg 7, D-55099 Mainz, Germany}
\affiliation{Department of Physics, Marquette University, Milwaukee, Wisconsin 53201, USA}
\affiliation{Institut f{\"u}r Kernphysik, Universit{\"a}t M{\"u}nster, D-48149 M{\"u}nster, Germany}
\affiliation{Bartol Research Institute and Dept. of Physics and Astronomy, University of Delaware, Newark, Delaware 19716, USA}
\affiliation{Dept. of Physics, Yale University, New Haven, Connecticut 06520, USA}
\affiliation{Columbia Astrophysics and Nevis Laboratories, Columbia University, New York, New York 10027, USA}
\affiliation{Dept. of Physics, University of Oxford, Parks Road, Oxford OX1 3PU, England, United Kingdom}
\affiliation{Dipartimento di Fisica e Astronomia Galileo Galilei, Universit{\`a} Degli Studi di Padova, I-35122 Padova PD, Italy}
\affiliation{Dept. of Physics, Drexel University, 3141 Chestnut Street, Philadelphia, Pennsylvania 19104, USA}
\affiliation{Physics Department, South Dakota School of Mines and Technology, Rapid City, South Dakota 57701, USA}
\affiliation{Dept. of Physics, University of Wisconsin, River Falls, Wisconsin 54022, USA}
\affiliation{Dept. of Physics and Astronomy, University of Rochester, Rochester, New York 14627, USA}
\affiliation{Department of Physics and Astronomy, University of Utah, Salt Lake City, Utah 84112, USA}
\affiliation{Dept. of Physics, Chung-Ang University, Seoul 06974, Republic of Korea}
\affiliation{Oskar Klein Centre and Dept. of Physics, Stockholm University, SE-10691 Stockholm, Sweden}
\affiliation{Dept. of Physics and Astronomy, Stony Brook University, Stony Brook, New York 11794-3800, USA}
\affiliation{Dept. of Physics, Sungkyunkwan University, Suwon 16419, Republic of Korea}
\affiliation{Institute of Basic Science, Sungkyunkwan University, Suwon 16419, Republic of Korea}
\affiliation{Institute of Physics, Academia Sinica, Taipei, 11529, Taiwan}
\affiliation{Dept. of Physics and Astronomy, University of Alabama, Tuscaloosa, Alabama 35487, USA}
\affiliation{Dept. of Astronomy and Astrophysics, Pennsylvania State University, University Park, Pennsylvania 16802, USA}
\affiliation{Dept. of Physics, Pennsylvania State University, University Park, Pennsylvania 16802, USA}
\affiliation{Dept. of Physics and Astronomy, Uppsala University, Box 516, SE-75120 Uppsala, Sweden}
\affiliation{Dept. of Physics, University of Wuppertal, D-42119 Wuppertal, Germany}
\affiliation{Deutsches Elektronen-Synchrotron DESY, Platanenallee 6, D-15738 Zeuthen, Germany}

\author{R. Abbasi}
\affiliation{Department of Physics, Loyola University Chicago, Chicago, Illinois 60660, USA}
\author{M. Ackermann}
\affiliation{Deutsches Elektronen-Synchrotron DESY, Platanenallee 6, D-15738 Zeuthen, Germany}
\author{J. Adams}
\affiliation{Dept. of Physics and Astronomy, University of Canterbury, Private Bag 4800, Christchurch, New Zealand}
\author{S. K. Agarwalla}
\thanks{also at Institute of Physics, Sachivalaya Marg, Sainik School Post, Bhubaneswar 751005, India.}
\affiliation{Dept. of Physics and Wisconsin IceCube Particle Astrophysics Center, University of Wisconsin{\textemdash}Madison, Madison, Wisconsin 53706, USA}
\author{J. A. Aguilar}
\affiliation{Universit{\'e} Libre de Bruxelles, Science Faculty CP230, B-1050 Brussels, Belgium}
\author{M. Ahlers}
\affiliation{Niels Bohr Institute, University of Copenhagen, DK-2100 Copenhagen, Denmark}
\author{J.M. Alameddine}
\affiliation{Dept. of Physics, TU Dortmund University, D-44221 Dortmund, Germany}
\author{N. M. Amin}
\affiliation{Bartol Research Institute and Dept. of Physics and Astronomy, University of Delaware, Newark, Delaware 19716, USA}
\author{K. Andeen}
\affiliation{Department of Physics, Marquette University, Milwaukee, Wisconsin 53201, USA}
\author{C. Arg{\"u}elles}
\affiliation{Department of Physics and Laboratory for Particle Physics and Cosmology, Harvard University, Cambridge, Massachusetts 02138, USA}
% ...

\collaboration{IceCube Collaboration}
\thanks{Contact author: \href{mailto:analysis@icecube.wisc.edu}{analysis@icecube.wisc.edu}}
\author{V. Aushev}
\affiliation{Department of Nuclear and High Energy Physics, Taras Shevchenko National University of Kyiv, 01601 Kyiv, Ukraine}

\date{Received 18 February 2025; accepted 29 April 2025; published 3 June 2025}
\begin{abstract}

We report a study of the inelasticity distribution in the scattering of neutrinos of energy $80-560~\mathrm{GeV}$ off nucleons. Using atmospheric muon neutrinos detected in IceCube's sub-array DeepCore during 2012--2021, we fit the observed inelasticity in the data to a parameterized expectation and extract the values that describe it best. Finally, we compare the results to predictions from various combinations of perturbative QCD calculations and atmospheric neutrino flux models.
\\ \\
\noindent
DOI: \href{https://doi.org/10.1103/PhysRevD.111.112001}{10.1103/PhysRevD.111.112001}

\end{abstract}

\maketitle
\renewcommand*{\thefootnote}{\fnsymbol{footnote}}
\setcounter{footnote}{3}
\renewcommand*{\thefootnote}{\arabic{footnote}}
\setcounter{footnote}{0}
\let\thefootnote\relax\footnotetext{\\{\it Published by the American Physical Society under the terms of
the \href{https://creativecommons.org/licenses/by/4.0/}{Creative Commons Attribution 4.0 International} license.
Further distribution of this work must maintain attribution to
the author(s) and the published article’s title, journal citation,
and DOI. Funded by SCOAP$^3$.}}

\newcommand{\numucc}{$\barparena\nu_{\mu}$~CC~}
\newcommand{\numu}{$\barparena\nu_{\mu}$~}
\newcommand{\numueq}{\barparena\nu_{\mu}}
\newcommand{\nue}{$\barparena\nu_e$~}
\newcommand{\nueqe}{\barparena\nu_e}

\newcommand{\coszeneq}{\cos{\theta_{z}}}

\newcommand{\Ereco}{$E_{\mathrm{reco}}$~}
\newcommand{\yreco}{$y_{\mathrm{reco}}$~}
\newcommand{\Etrue}{$E_{true}$~}
\newcommand{\ytrue}{$y_{true}$~}

\newcommand{\meany}{$\langle y\rangle$~}
\newcommand{\meanyeq}{\langle y\rangle}

\newcommand{\lglam}{$\log_{10}\lambda$~}

\newcommand{\nunucl}{$\nu-n$~}

\newcommand{\nubar}{\bar{\nu}}

\newcommand{\nurat}{$\nu/\nubar$~}
\newcommand{\dpdy}{$\frac{dp}{dy}$~}

\section{Introduction~\label{sec:intro}}
Neutrinos interact weakly, and are therefore unique probes to test our understanding of the Standard Model of particle physics.
Until recently, neutrino interactions have been studied at accelerators from tens of MeV to 340~GeV~\cite{COHERENT:2020iec,eV2EeV}; however, two new detectors placed near Collision Point~1 at the Large Hadron Collider (LHC) have started extending this range. The FASER$\nu$ Collaboration has performed a first measurement of the average total $\nu_{e}$ ($\nu_{\mu}$) cross section for neutrinos of energy $560-1740$~GeV ($520-1760$~GeV)~\cite{FASER:2024hoe}, while SND@LHC has reported their first neutrino candidates~\cite{Ilieva:2024vpu}.

Another source of information on neutrino interactions are neutrino telescopes which detect naturally created neutrinos. These experiments access much higher energies, but are limited by the uncertainties on the incoming flux and the accuracy with which an interaction can be reconstructed. Despite these limitations, these experiments can contribute in a complimentary way to the understanding of neutrino cross sections. IceCube, which detects both atmospheric and astrophysical neutrinos, has previously reported measurements of the total neutrino cross section up to a PeV~\cite{IceCube:2020rnc,IceCube_xsec2:2017roe} as well as a measurement of the inelasticity distribution at neutrino energies of $1 - 20$~TeV~\cite{IC_inelasticity_paper}. Such energies are well above those accessible at terrestrial accelerators.

Here, we present the results from a measurement of the flux-averaged inelasticity distribution in neutrino-nucleon interactions from 80 to 560~GeV using data from the IceCube DeepCore array. At these energies neutrinos interact mainly through deep-inelastic scattering (DIS), which involves scattering off a quark or gluon in the nucleon, resulting in a lepton and a hadronic shower in the final state. The inelasticity $y$ is a measure of the energy distribution of the products of the interaction, defined as~\cite{Devenish:2004pb}
\begin{equation}
    y = 1 - E_{\mathrm{lepton}}/E_\nu,
\end{equation}
where $E_\nu$ is the energy of the incoming neutrino and $E_{\mathrm{lepton}}$ is the energy of the outgoing lepton. 
In this study, we follow Ref.~\cite{IC_inelasticity_paper} and fit an empirical model of the inelasticity distribution shape to atmospheric neutrino data at different values of $E_\nu$, rather than doing a differential cross section measurement.

The inelasticity distribution of neutrino-nucleon interactions simultaneously probes the fundamental properties of the weak force and our understanding of the structure of the nucleon in QCD. 
At energies below 1~PeV neutrinos and antineutrinos have different inelasticity distributions as the bulk of their interactions happen on valence quarks. The difference disappears at high energies as neutrinos and antineutrinos are able to interact with equal numbers of sea quarks and antiquarks and with gluons~\cite{Devenish:2004pb}.
The result presented here uses the atmospheric neutrino flux, which contains both neutrinos and antineutrinos. We measure a convolution of the neutrino and antineutrino inelasticity distributions with the neutrino to antineutrino ratio. 
Such a study can validate current cross section and flux models, while any deviation in the extracted inelasticity distribution could indicate a mismodeling either in cross section or in atmospheric flux predictions.
We also invert our results to extract the neutrino fraction in our sample, assuming a specific cross section calculation.

At $E_\nu=$~80--560~GeV, neutrino-nucleon interactions are expected to be dominated by DIS; however, resonance (RES) and quasielastic (QE) scattering events provide a non-negligible contribution at low inelasticity values. In this study, we report the inclusive inelasticity distribution. This is because in DeepCore we cannot reliably differentiate between QE, RES, and DIS events due to the large intermodule spacing compared to the characteristic size of the hadronic cascades.

We begin by introducing the IceCube Neutrino Observatory (Sec.~\ref{sec:detector}), followed by a description of the data sample used (Sec.~\ref{sec:selection}). The analysis is presented in Sec.~\ref{sec:analysis}, and results are presented in Sec.~\ref{sec:results}. In Sec.~\ref{sec:discussion}, we discuss the implications of our results.
The limitations of the parametrization used in our analysis are described in detail in Appendix~\ref{apx:param_lim}. In Appendix~\ref{apx:param_impact}, we show a detailed comparison of the impact of measured and nuisance parameters on our data sample.

\section{IceCube neutrino detector~\label{sec:detector}}
The IceCube Neutrino Observatory is an ice-Cherenkov detector consisting of 5160 digital optical modules (DOMs) deployed deep under the surface at the geographic South Pole. 
The DOMs instrument a volume of approximately 1~km$^3$, starting at a depth of 1450~m below the surface and ending at 2450~m. The bulk of the array is optimized for the observation of neutrinos with energies of 1~TeV and above and has DOMs spaced by 17~m in the vertical direction and 125~m in the horizontal direction~\cite{IceCube_instrumentation:2016zyt}. A low-energy extension, known as DeepCore, is located at the bottom, central region of the detector, where higher quantum efficiency photomultiplier tubes (PMTs) are deployed closer together, with a vertical separation of 7~m and a horizontal separation between 40 and 70~m~\cite{IceCube_deepcore:2011ucd}. DeepCore can observe neutrino events down to $\sim$10~GeV~\cite{IceCubeCollaboration:2023wtb}.

The sensitive element in the DOMs is a 10~in. PMT facing downward away from the ice surface~\cite{IceCube_DOM:Hanson:2006bk}. The PMT is enclosed in a glass sphere together with the electronics for power and signal digitization~\cite{IceCube_instrumentation:2016zyt}. 
The waveforms produced by the PMTs in each DOM are recorded by analog-digital converters, and the digitized data are passed to a pulse-finding algorithm, which deconvolves it into pulses using the typical response of the PMT to single photons~\cite{IceCube_reco_old:2013dkx,IceCube_spe_calibration:2020nwx}. The interpretation of the properties of the events, such as the event selection and reconstruction described below, is carried out based on the information contained in these pulses.

In this analysis, we use DeepCore to study the inelasticity of events with primary neutrino energies in the range from $80$ to $560$~GeV. The hadronic cascade in such events is still mostly contained inside the more densely instrumented DeepCore volume, which gives us a more precise reconstruction compared to the main IceCube array. 
The energy of the muons exiting the DeepCore, but still contained inside the IceCube, can be estimated from their track length.

\section{Data selection~\label{sec:selection}}

The type of events that are most useful for this analysis are those where the outgoing lepton can be distinguished from hadronic interaction products, and the inelasticity can therefore be estimated. This limits us to $\nu_\mu$-nucleon charged-current (CC) interactions, where the products are a hadronic shower starting at the interaction point and developing over a distance of meters, and a muon that travels roughly 4.5~m per GeV of its kinetic energy.

The data sample used for this analysis comes from the common DeepCore event selection described in Ref.~\cite{IceCubeCollaboration:2023wtb}, specifically at \textit{Level 5}. 
The common selection focuses on rejecting events produced by atmospheric muons as well as detector noise, while keeping events where the neutrino interacted inside the DeepCore region.
Based on Monte Carlo modeling, the sample has approximately a $2:1$~ratio of neutrinos to non-neutrino backgrounds after the common selection, so further levels of selection described in this section are necessary for refining a high purity $\nu_\mu$~CC-only sample for this study.

\subsection{Event reconstruction and classification}

Selected events are reconstructed using RETRO, an IceCube low-energy reconstruction method described in Ref.~\cite{IceCube_retro_reco:2022kff}. We reconstruct this sample under the hypothesis that the interaction produces a hadronic shower and a muon, each with a distinct energy, but traveling in the same direction. This hypothesis is reasonable for $\nu_\mu$~CC interactions with $E_\nu>10$~GeV, the events that are most useful for this study. The inelasticity can therefore be estimated as
\begin{equation}
    y_{\mathrm{reco}} = 1 - E_{l, \mathrm{reco}}/(E_{l, \mathrm{reco}} + E_{h, \mathrm{reco}}),
\end{equation}
where $reco$ stands for reconstructed, $E_l$ is the energy of the lepton, and $E_h$ is the total energy of the hadronic shower.

\begin{table}[h]
\centering
\begin{tabular}{lcc}
\hline
                                 & Events in 9.28~years~~ & \% of sample \\ \hline
  DIS $\nu_\mu$ CC              & 4756                 & 68.3~\%                \\ 
 DIS $\bar{\nu}_\mu$ CC           & 1965                 & 28.0~\%                \\
  Non-DIS $\nu_\mu + \bar{\nu}_\mu$ CC           & 188                 & 2.7~\%                \\\hline
 $\nu$ NC                   & 27                   & 0.4~\%                 \\ 
 $\nu_e+\bar{\nu}_e$           & 20                   & 0.3~\%                 \\ 
 Atm. $\mu$              & 11                   & 0.2~\%                 \\
 $\nu_\tau+\bar{\nu}_\tau$      & 10                    & 0.1~\%                 \\ 
 Noise                   & 0                    & 0.0~\%                 \\ \hline
\end{tabular}
\caption{Sample composition at the final level of selection, as described in Sec.~\ref{sec:selection}, with projected event counts for 9.28 years of data, calculated using the baseline Monte Carlo set.}
\label{tab:final_sample}
\end{table}

\begin{figure}
    \centering
    \includegraphics[width=0.95\linewidth]{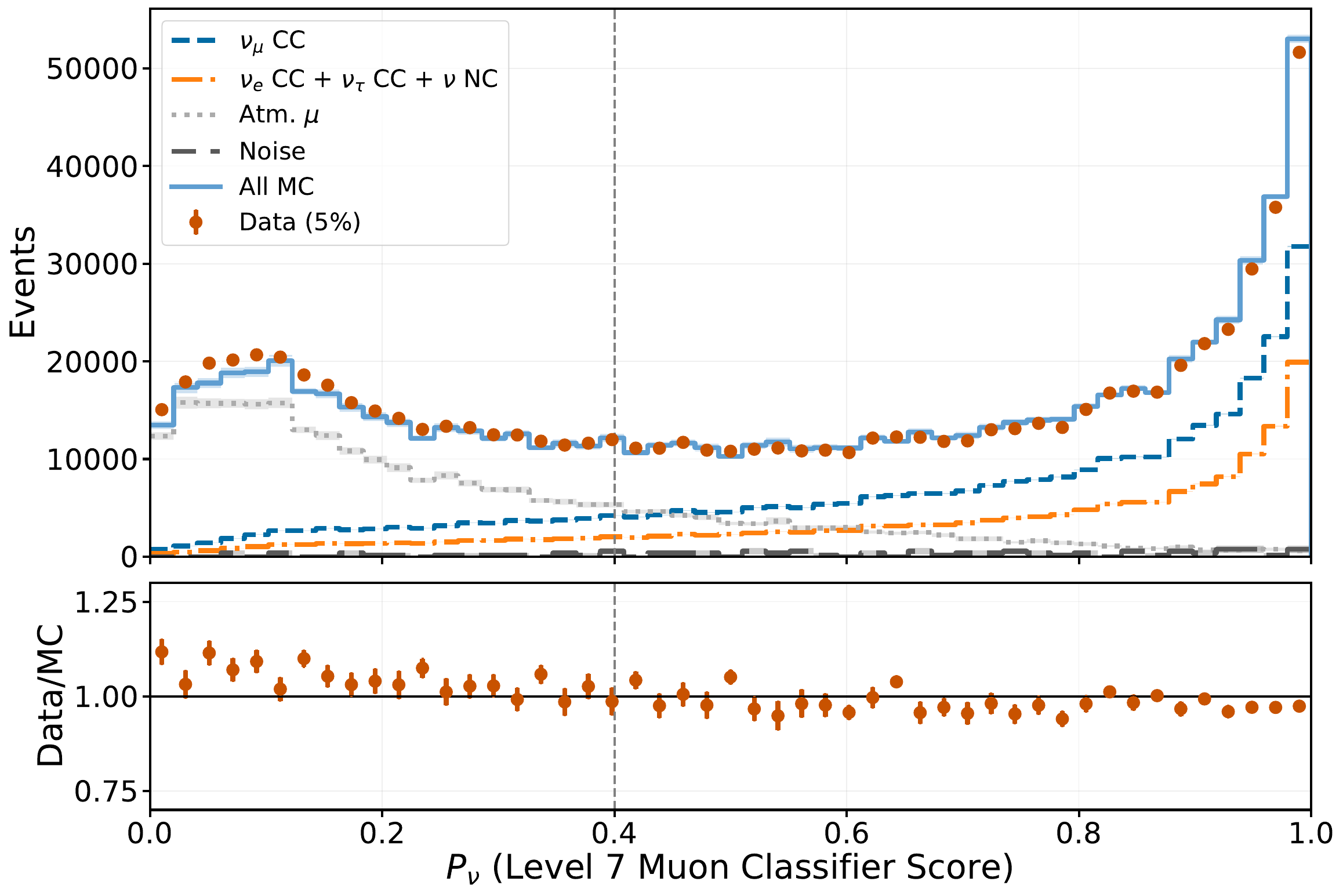}
    \caption{Muon Classifier score comparison for $5\%$ of data and for predicted sample composition at the reconstruction level. The Monte Carlo normalization is scaled to match total event counts to data. The gray dashed line indicates the position of the cut on Muon Classifier score, with values to the left of the line removed from the final sample. Tinted bands around histogram lines represent $1\sigma$ statistical uncertainty.}
    \label{fig::sel::mu_classif_kayla}
\end{figure}

After the reconstruction is performed, two boosted decision tree (BDT) classifiers are applied to the events. Both classifiers were trained on simulation and rely on reconstructed variables~\cite{LeonardDeHolton:2022tyg}. The first is Muon Classifier, which is trained to identify remaining atmospheric muons in the sample. It uses inputs of five variables, namely the Level 4 Muon Classifier score~\cite{IceCubeCollaboration:2023wtb}, vertical and radial position of the reconstructed interaction vertex, vertical location of the deepest pulse recorded in the ice which matches a hypothesis of muon traveling in a ``corridor'' between strings, and the difference between reconstructed directions using RETRO and SANTA reconstruction algorithms~\cite{IceCube_retro_reco:2022kff}. A comparison of Muon Classifier score for data and simulation is shown on Fig.~\ref{fig::sel::mu_classif_kayla}.

\begin{figure}
    \centering
    \includegraphics[width=0.95\linewidth]{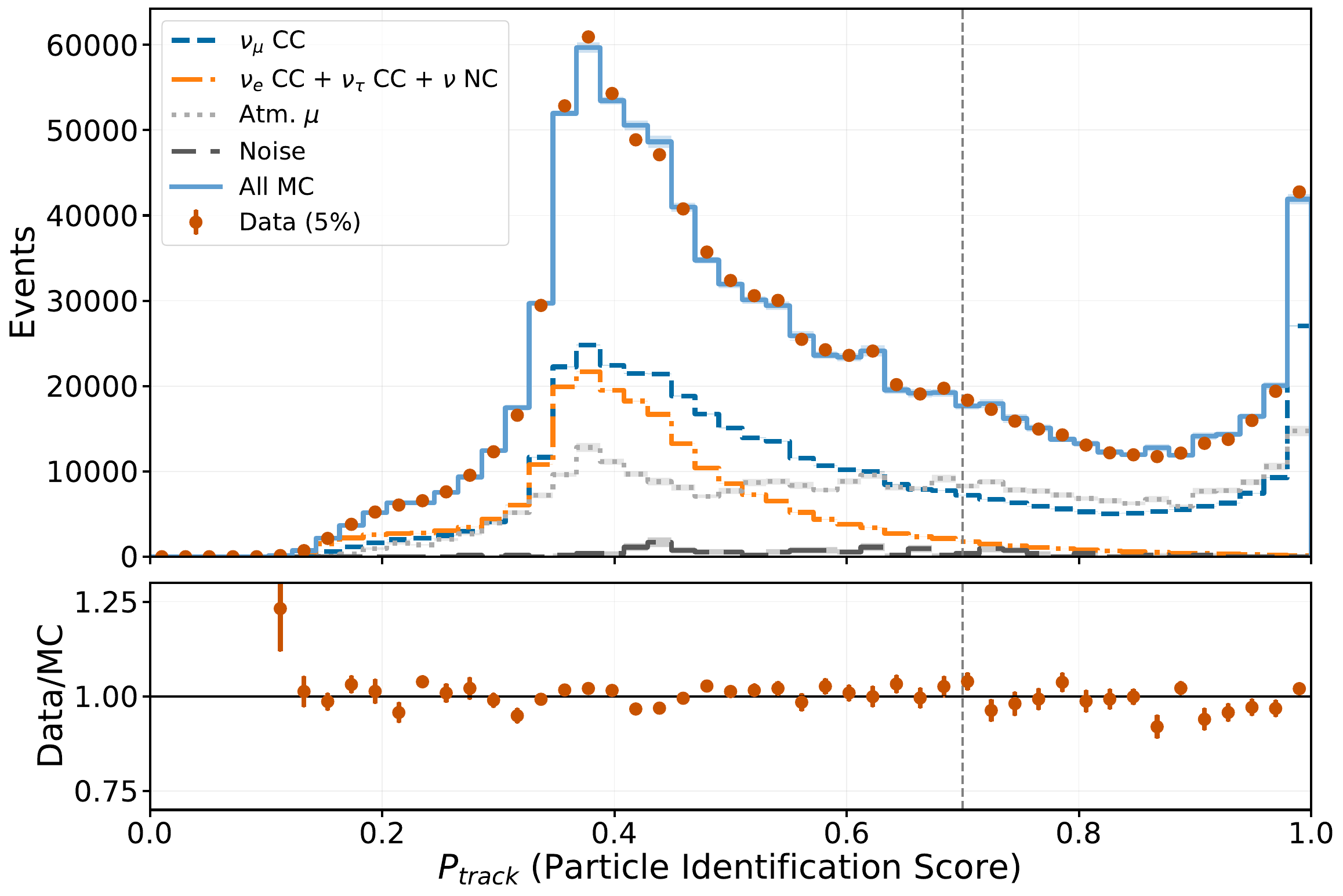}
    \caption{PID score comparison for $5\%$ of data and for predicted sample composition at the reconstruction level. The Monte Carlo normalization is scaled to match total event counts to data. The gray dashed line indicates the position of the cut on PID score, with values to the left of the line removed from the final sample. Tinted bands around histogram lines represent $1\sigma$ statistical uncertainty.}
    \label{fig::sel::pid_classif_kayla}
\end{figure}

The second BDT classifier determines the Particle Identification Score (PID), which divides events into tracklike and cascadelike. The classifier is trained to identify $\nu_\mu$~CC interactions as tracklike events, while $\nu_e$~CC interactions represent cascadelike events.
The BDT uses input of five variables coming from the reconstruction algorithm: track length of the outgoing muon, cascade energy, zenith angle, zenith angle uncertainty, and the likelihood difference between the best-fit hypothesis with arbitrarily long track and with track length fixed at zero. A comparison of PID classifier score for data and simulation is shown on Fig.~\ref{fig::sel::pid_classif_kayla}, which also indicates the cut on PID~score~$>$~0.7 described in Sec.~\ref{sec:selection:non_numucc_bkg}.

\begin{figure}
    \centering
    \includegraphics[width=0.95\linewidth]{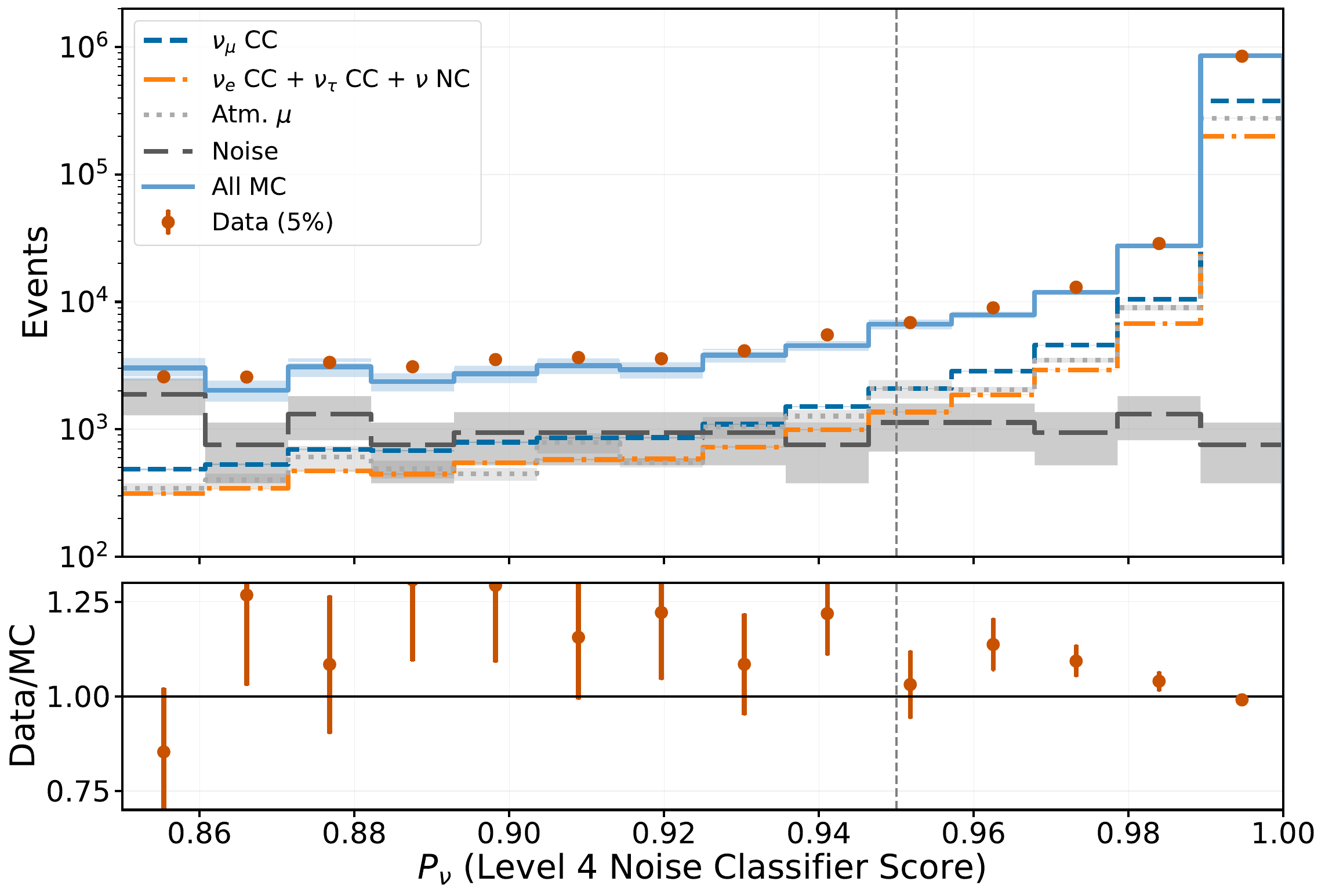}
    \caption{Level 4 Noise Classifier score comparison for $5\%$ of data and for predicted sample composition at the reconstruction level. The Monte Carlo normalization is scaled to match total event counts to data. The gray dashed line indicates the position of the additional cut on the classifier score, with values to the left of the line removed from the final sample. Tinted bands around histogram lines represent $1\sigma$ statistical uncertainty.}
    \label{fig::sel::noise_classif_L_four}
\end{figure}

\begin{figure*}
    \centering
    \includegraphics[width=0.45\linewidth]{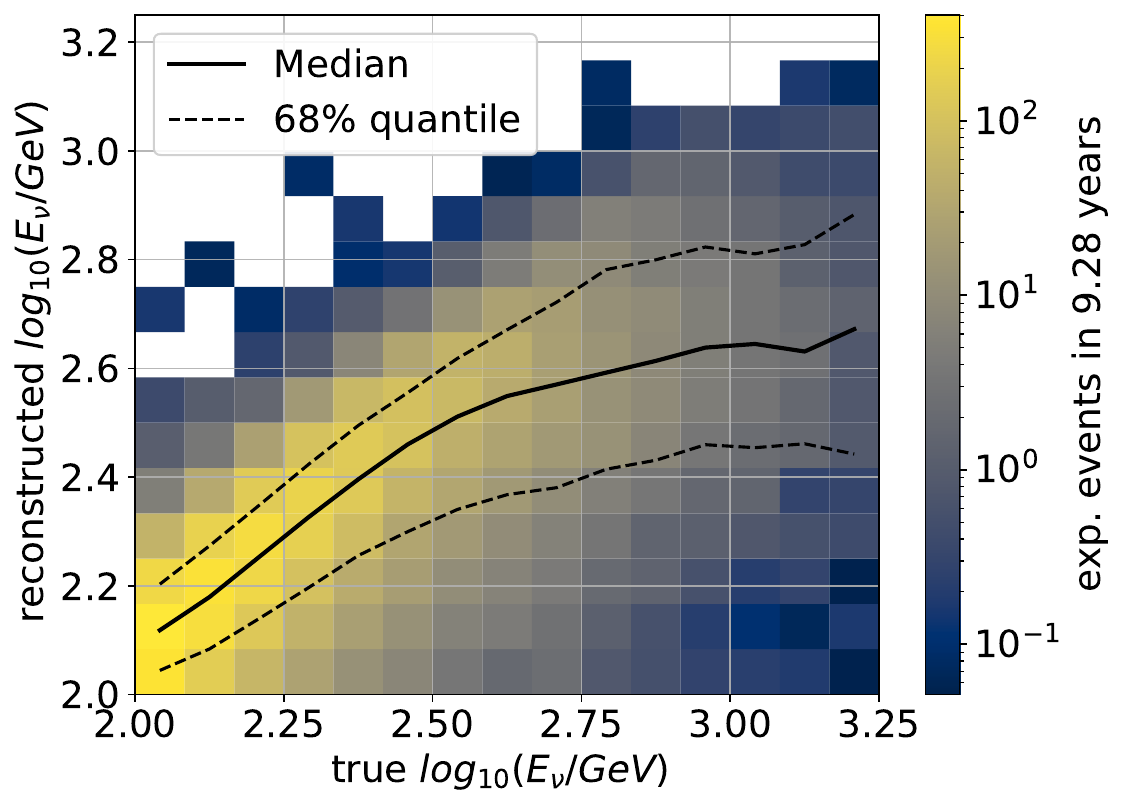}
    \includegraphics[width=0.45\linewidth]{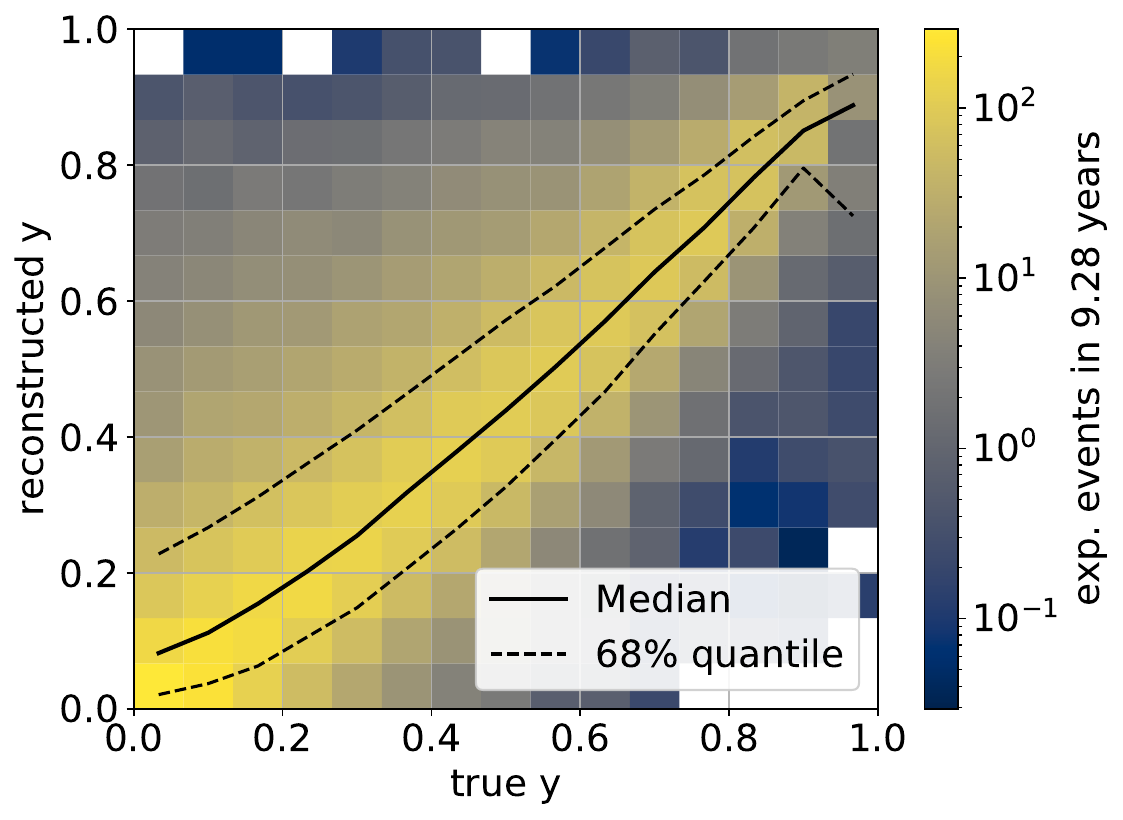}
    \caption{Reconstruction resolutions for neutrino energy (left) and inelasticity (right) at the final level of event selection.}
    \label{fig:reco_res}
\end{figure*}

To further reduce non-neutrino background, the Muon Classifier score is required to be $> 0.4$ for events to pass to the next step of selection. An additional cut of $>0.95$ is applied on the Level 4 Noise Classifier~\cite{IceCubeCollaboration:2023wtb} to reduce remaining background. The performance of the Level 4 Noise Classifier score for our sample at the reconstruction level is shown in Fig.~\ref{fig::sel::noise_classif_L_four}.

\subsection{Rejection of non-neutrino backgrounds}

Another class of variables used to remove noise and low-quality events are calculated based on estimated photon speeds, under the assumption that each detected photon has started at the event interaction vertex and then traveled without scattering to the corresponding DOM.
Photon speeds estimated in this way will not necessarily accurately represent the real event; however, the variables calculated using these are still useful, as they provide information on the event quality. In our event selection, we require the median estimated speed of all detected photons in an event to be $>0.4c$, where $c$ is the speed of light in vacuum. This helps us reject noise-induced events which may to have lower median speeds after being reconstructed with physics event hypothesis. We also require that in each event the fraction of detected photons with estimated speeds below zero or above $c$ is $<0.4$ because such pulses typically result from the PMT noise. Both of the thresholds were chosen based on Monte Carlo modeling. A photon speed below zero indicates that the pulse happened before the reconstructed interaction time, while photon speed above $c$ means that the pulse happened either too far from the reconstructed vertex or too early to be explained by a photon traveling from the vertex at the speed $c$. While a photon traveling in ice will have a speed of $c/n$, where $n$ is the refractive index of ice, the choice was made nevertheless to make a cut at the speed of $c$, to avoid penalizing events where a muon travels at speed higher than $c/n$ and then produces a Cherenkov photon near the DOM.

Restrictions are also applied to reject neutrino events with likely coincident atmospheric muons entering the detector. This is achieved by introducing cuts on events with pulses in the veto region. This region includes the top and the outermost optical modules of IceCube array, as these are more likely to detect light from atmospheric muons entering the detector. We limit the allowed number of pulses in the outermost layer of IceCube to be below 8. At the same time, we also require that there is no downward traveling trend in the pulse distribution in the top 15 layers of DOMs in IceCube, which corresponds to a vertical range of roughly 255~m.

We introduce vertex containment cuts on reconstructed vertex position to remove events with vertex located outside the DeepCore. We require that the vertical position of the vertex be in the range -500~$< z_{\mathrm{reco}}<$~-200~m from the vertical center of IceCube and that horizontal radial distance from the central string be $\rho_{\mathrm{reco}}<$~300~m.

\subsection{Rejection of non-$\nu_\mu$~CC backgrounds}
\label{sec:selection:non_numucc_bkg}

The remaining steps focus on selecting $\nu_\mu$ CC events with high-quality reconstruction. The primary event types targeted for the removal are very \textit{dim} atmospheric muons, which leave few or no pulses on the veto region, and were therefore missed by previous selection steps, and non-$\nu_\mu$ CC neutrino interactions.

Dim muons are rejected by estimating how well the track hypothesis used fits the data. To do this, we require that at least 10~pulses should be observed within a time window of [-10,+900]~ns from their expected arrival time assuming Cherenkov emission without scattering. This cut removes an estimated 95\% of the muons, while reducing the signal by 20\%. 

Other neutrino interactions are rejected by applying a strict cut on the PID score, which is required to be higher that $0.7$ for the events to pass the selection. 
Events with $y_{\mathrm{reco}}<0.001$ or $E_{h,\mathrm{reco}} = 0$~GeV are removed, as they were found to contain a large fraction of events with the cascade located either in the less densely instrumented region outside the DeepCore or fully below the detector. Finally, we also introduce a cut on the reconstructed zenith direction $\cos\theta_{z,\mathrm{reco}}<-0.2$ to keep only up-going events, which are likely to illuminate the PMT inside a few DOMs without the need for the light to scatter. Upgoing events are also less likely to contain misidentified atmospheric muons due to the shielding by the earth.

\subsection{Final level sample}

A Monte Carlo estimate of the composition of the final sample is given in Table~\ref{tab:final_sample}. Our selection results in a signal purity of 99\%. The average ratio of neutrinos to antineutrinos is 2.4, which reflects the properties of the atmospheric neutrino flux, the neutrino cross section, and event selection efficiency dependence on true inelasticity. The data do not provide sufficient information to distinguish between neutrinos and antineutrinos, so the analysis is done for the sum of both.

The reconstruction resolutions for neutrino energy and inelasticity in the final simulated sample are presented in Fig.~\ref{fig:reco_res}. The energy estimator shows a clear correlation with neutrino energy up to 500~GeV. Above that, the resolution degrades, because the typical size of the hadronic cascade approaches the size of DeepCore and a larger portion of the muon track could be located outside DeepCore. The inelasticity estimator also displays a correlation across all values, with deviations observed toward the edges, as is expected from one-sided errors in those regions. These two reconstructed quantities are used to extract the inelasticity distribution, as described in the next section.

\section{Analysis method~\label{sec:analysis}}
\subsection{General strategy}
\label{sec::gen_info}

\begin{figure}
    \centering
    \includegraphics[width=0.99\linewidth]{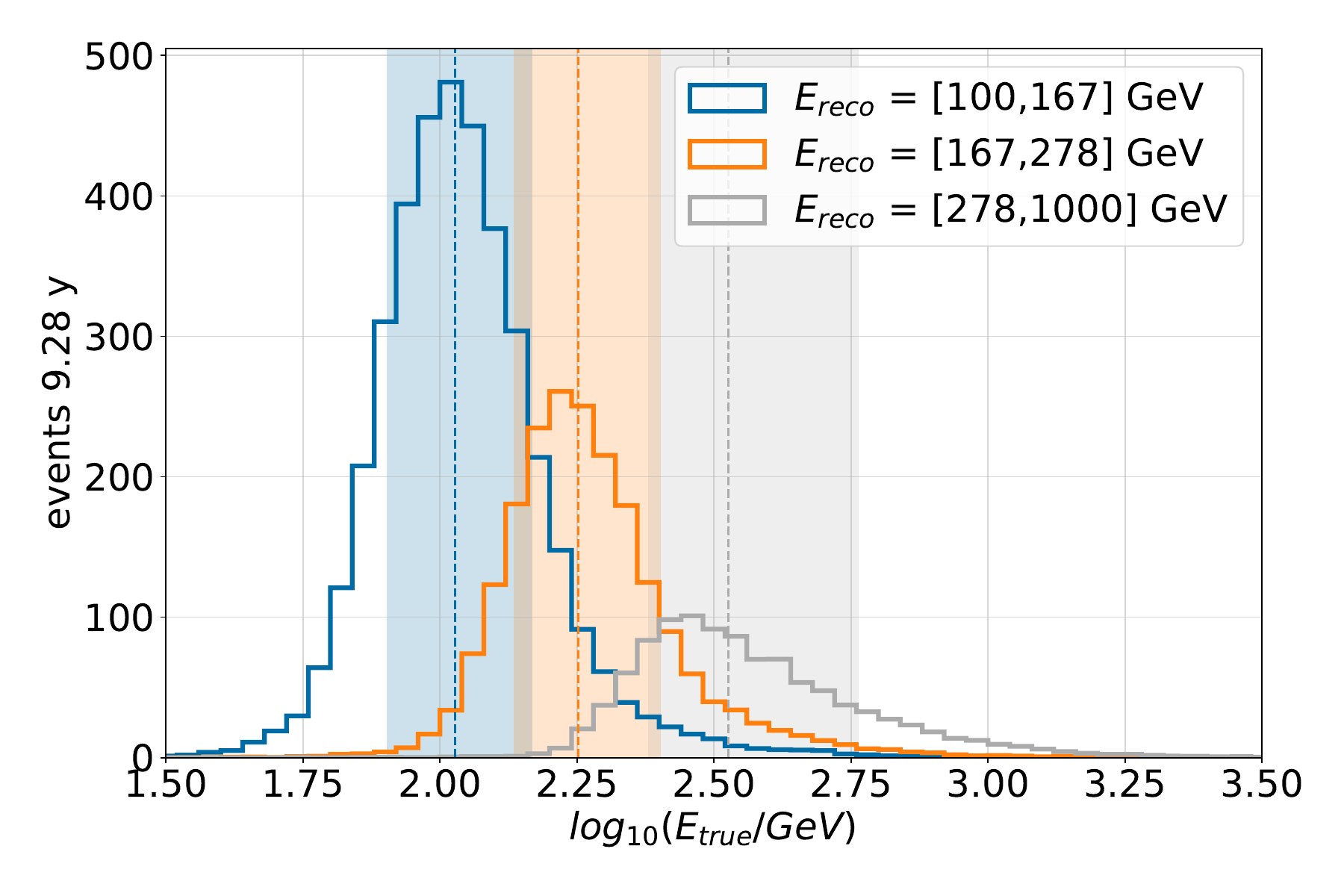}
    \caption{Predicted energy distributions for each reconstructed energy bin with 9.28 years of data using the {\sc genie} cross section model~\cite{Andreopoulos:2015wxa} and HKKMS'15 atmospheric neutrino flux~\cite{Honda:2015fha}. The shaded bands indicate the central 68\% quantile for the energy bin of a corresponding color and the dashed lines indicate the median true energy of the bin.}
    \label{fig:energy_binning}
\end{figure}

\begin{figure}
    \centering
    \includegraphics[width=0.99\linewidth]{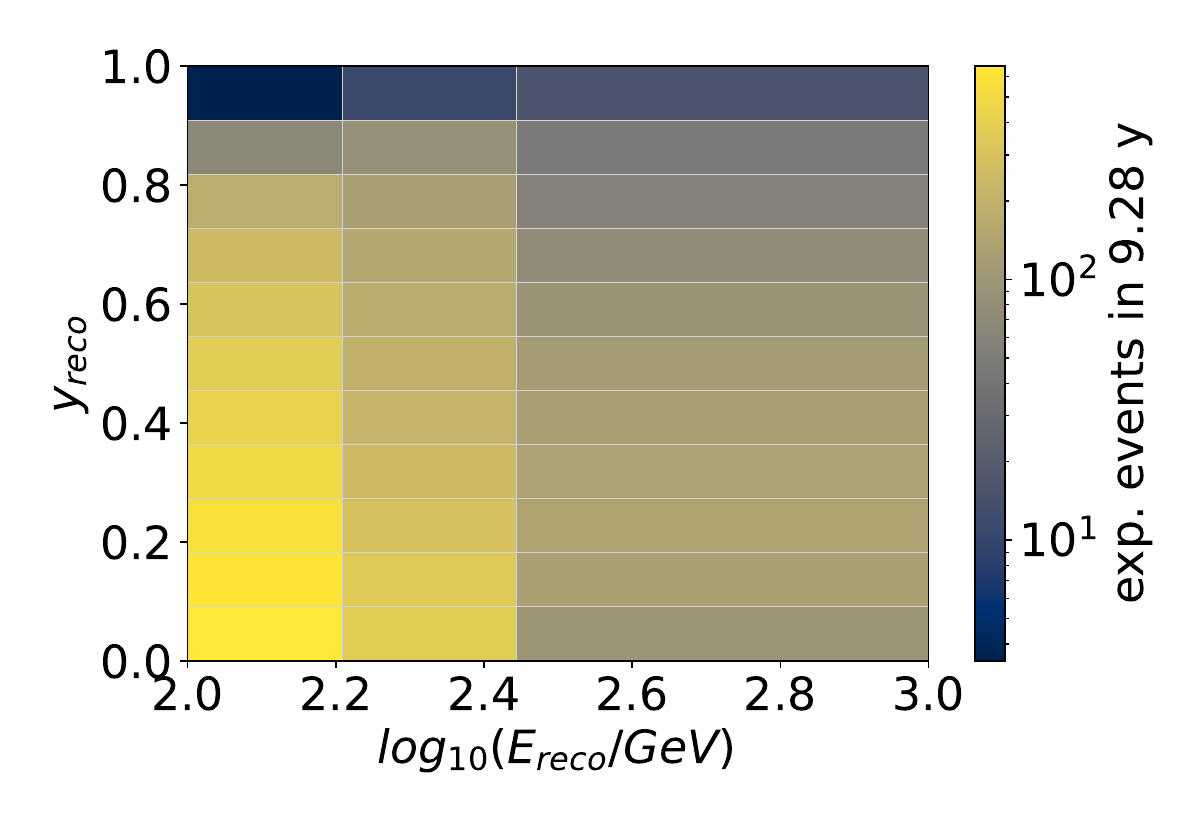}
    \caption{Expected event count distribution in the analysis template for 9.28 years of data, assuming the {\sc genie} cross section model and HKKMS’15 flux.}
    \label{fig:genie_template}
\end{figure}

\begin{figure*}
    \centering
    \includegraphics[width=0.35\linewidth]{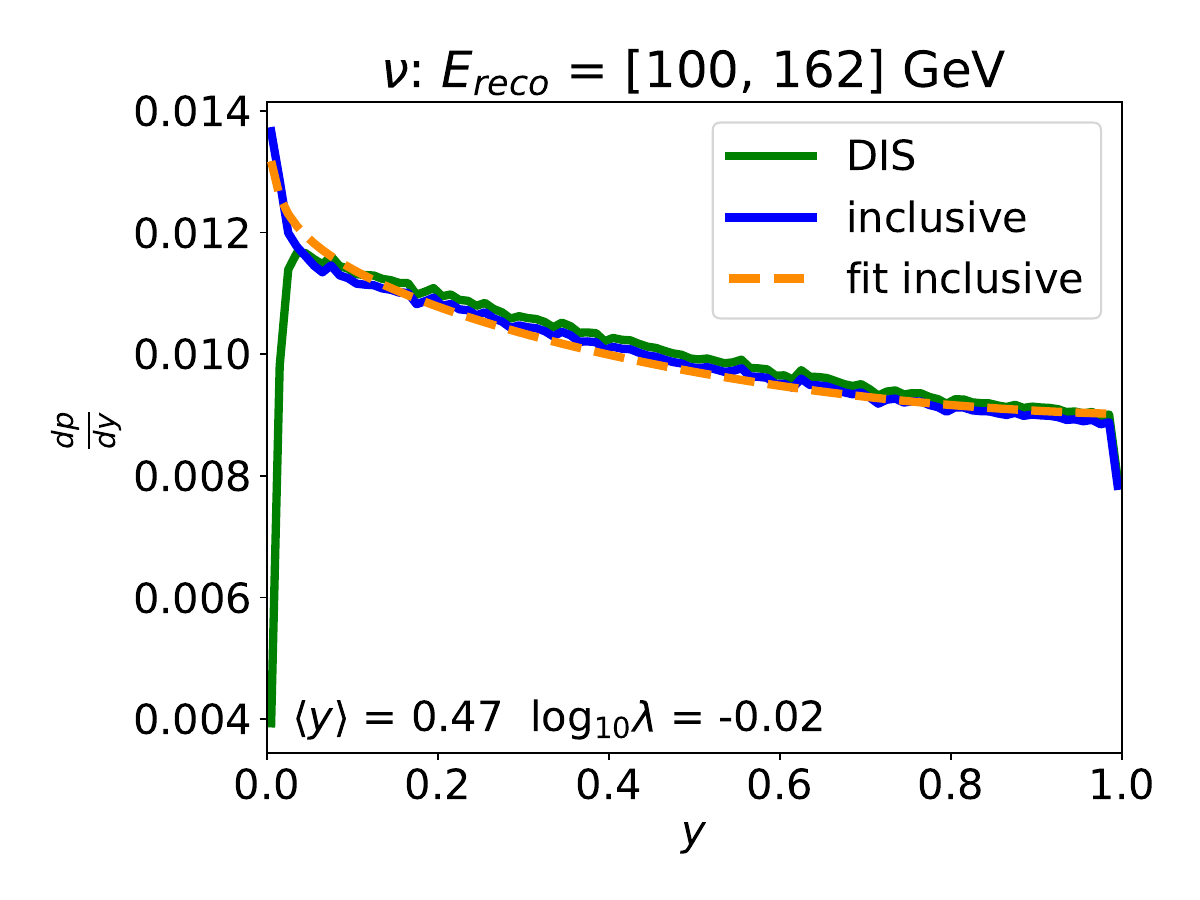}
    \includegraphics[width=0.35\linewidth]{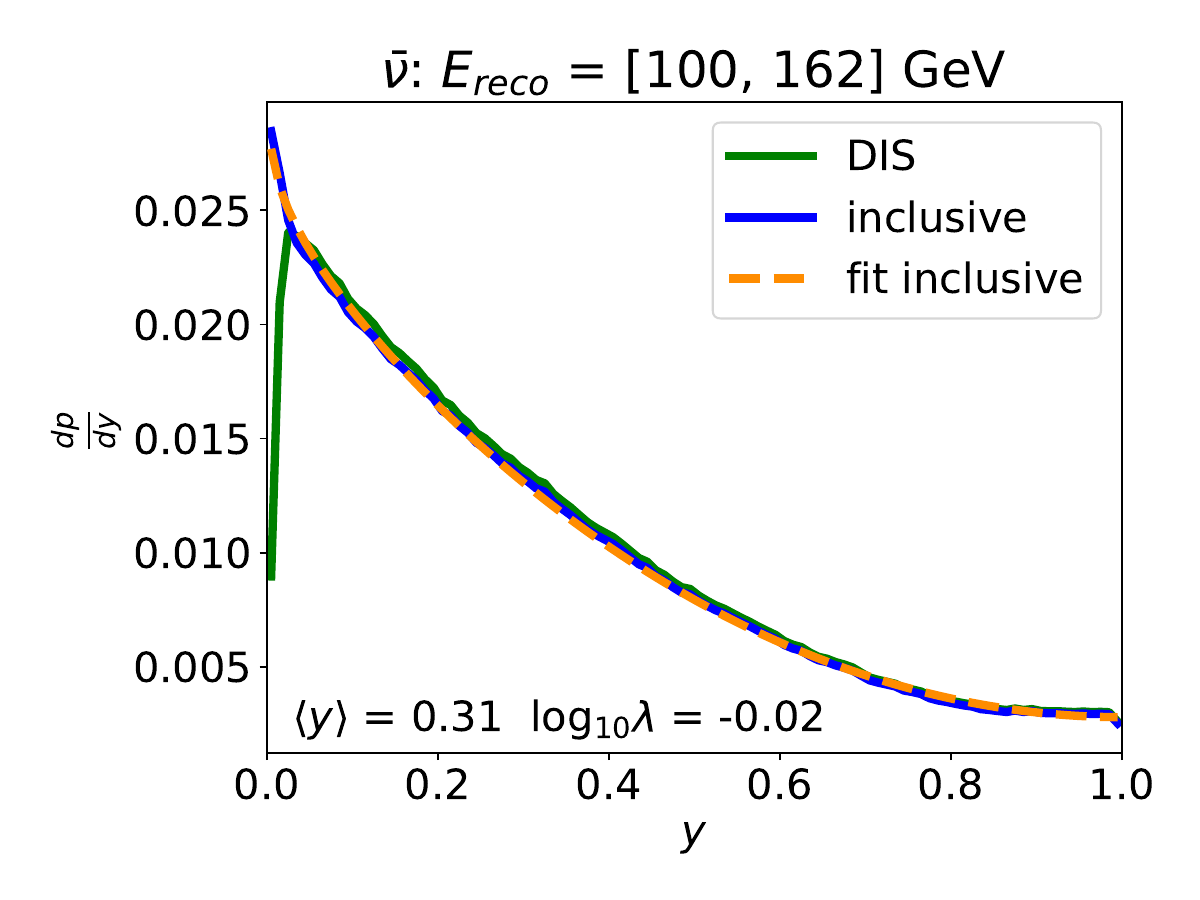}
    \includegraphics[width=0.35\linewidth]{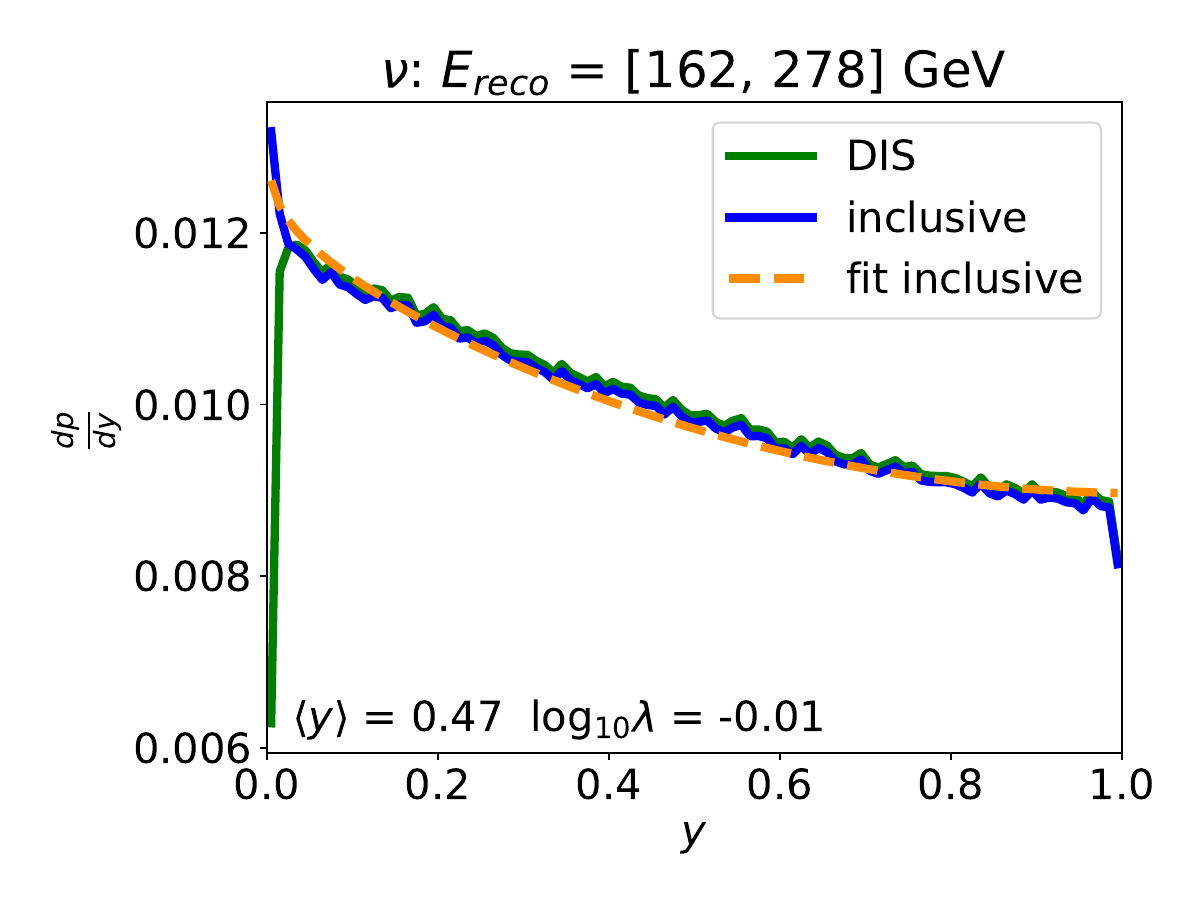}
    \includegraphics[width=0.35\linewidth]{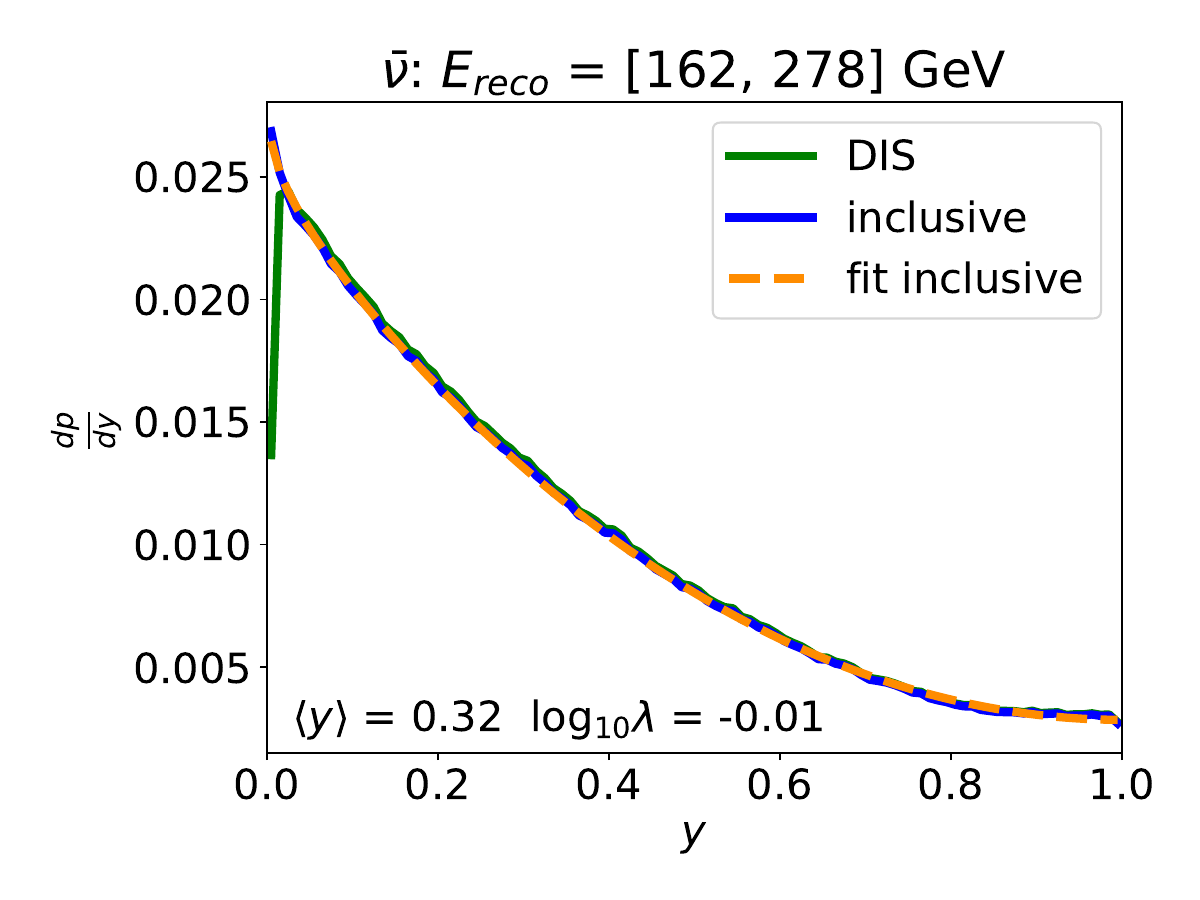}
    \includegraphics[width=0.35\linewidth]{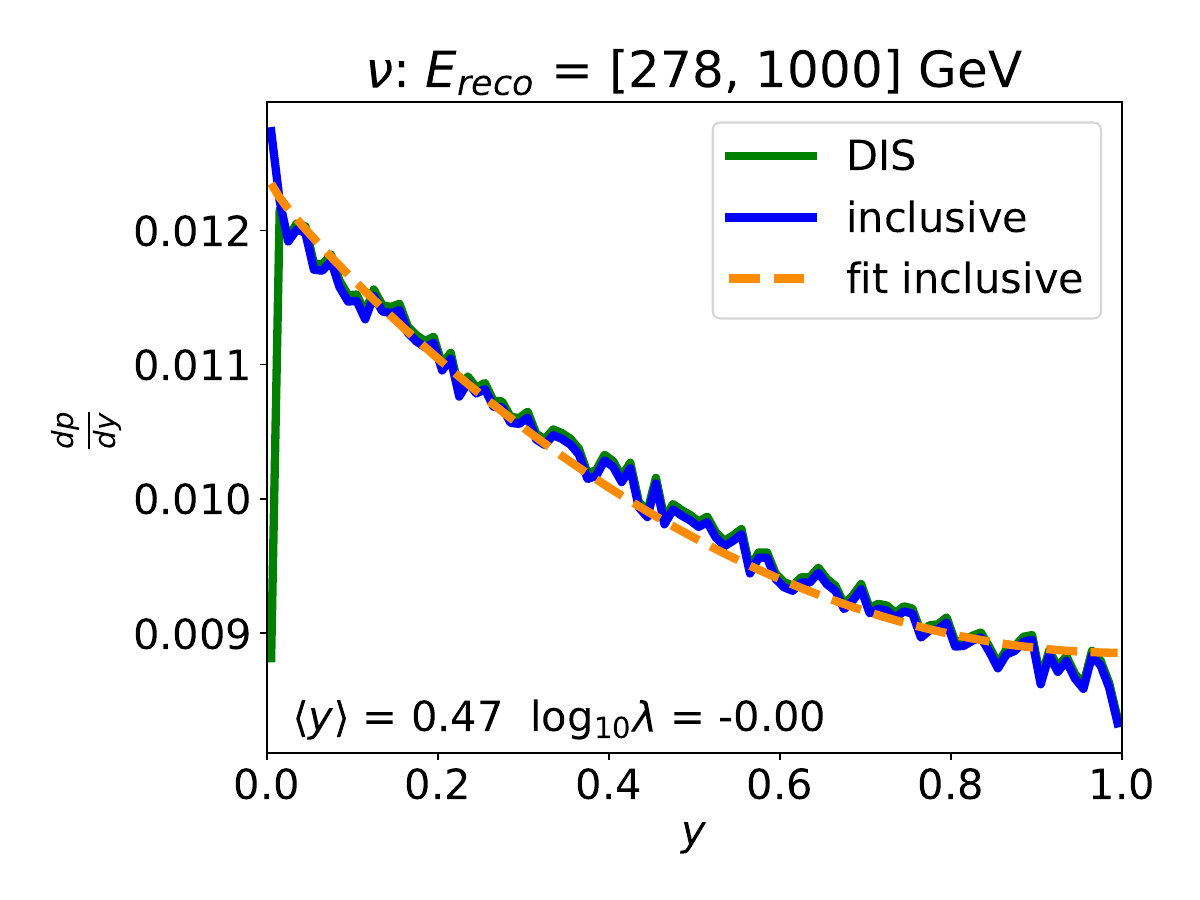}
    \includegraphics[width=0.35\linewidth]{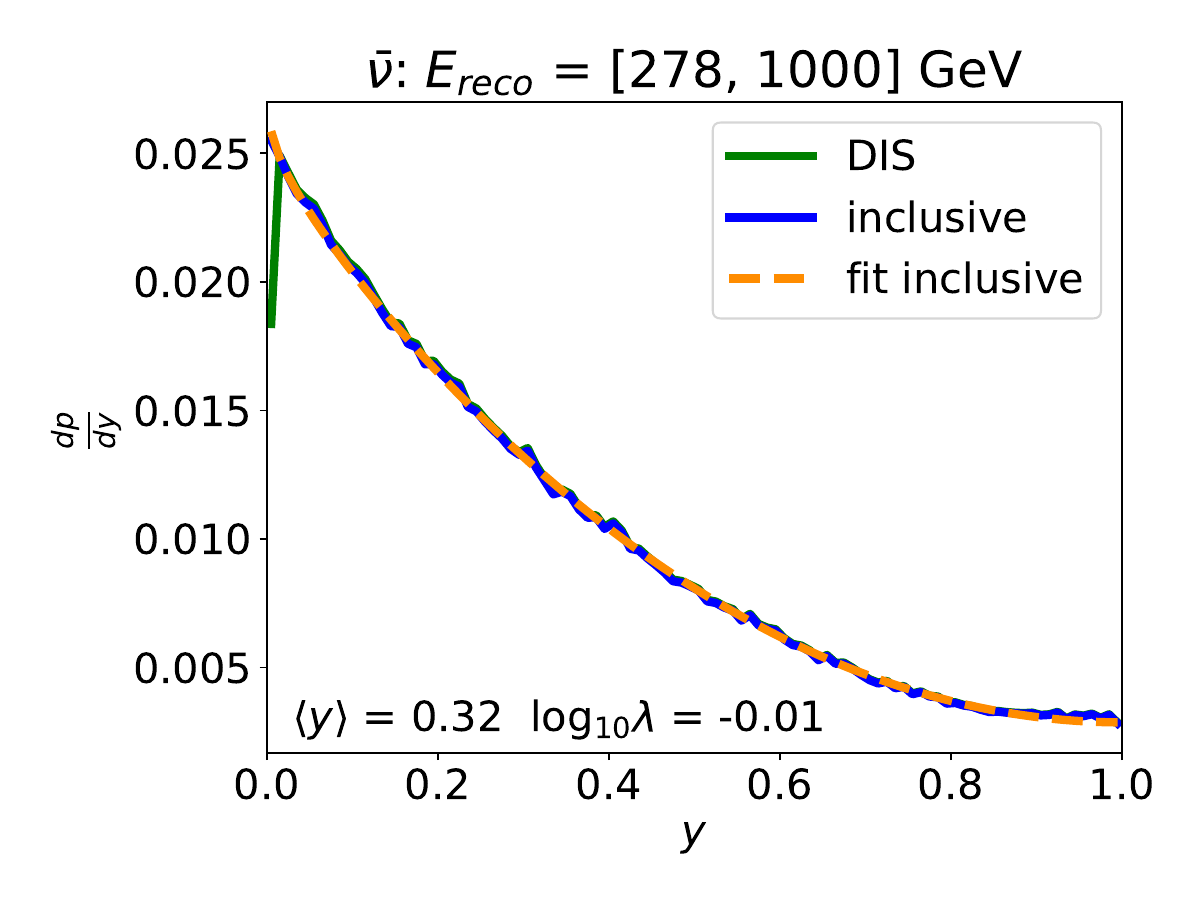}
    \caption{Fit of the parametrization (\ref{eq::param_dpdy}) to inclusive inelasticity distributions for generator level {\sc genie} Monte Carlo. Fits were done separately for neutrino and antineutrino inclusive inelasticity distributions in three energy bins, which were weighted to reflect energy distributions in energy bins at analysis level. Fit results for parameters \meany and \lglam are shown in the bottom left of each plot.}
    \label{fig::app::fit_to_step1_MC}
\end{figure*}

\begin{figure*}
    \centering
    \includegraphics[width=0.3\linewidth]{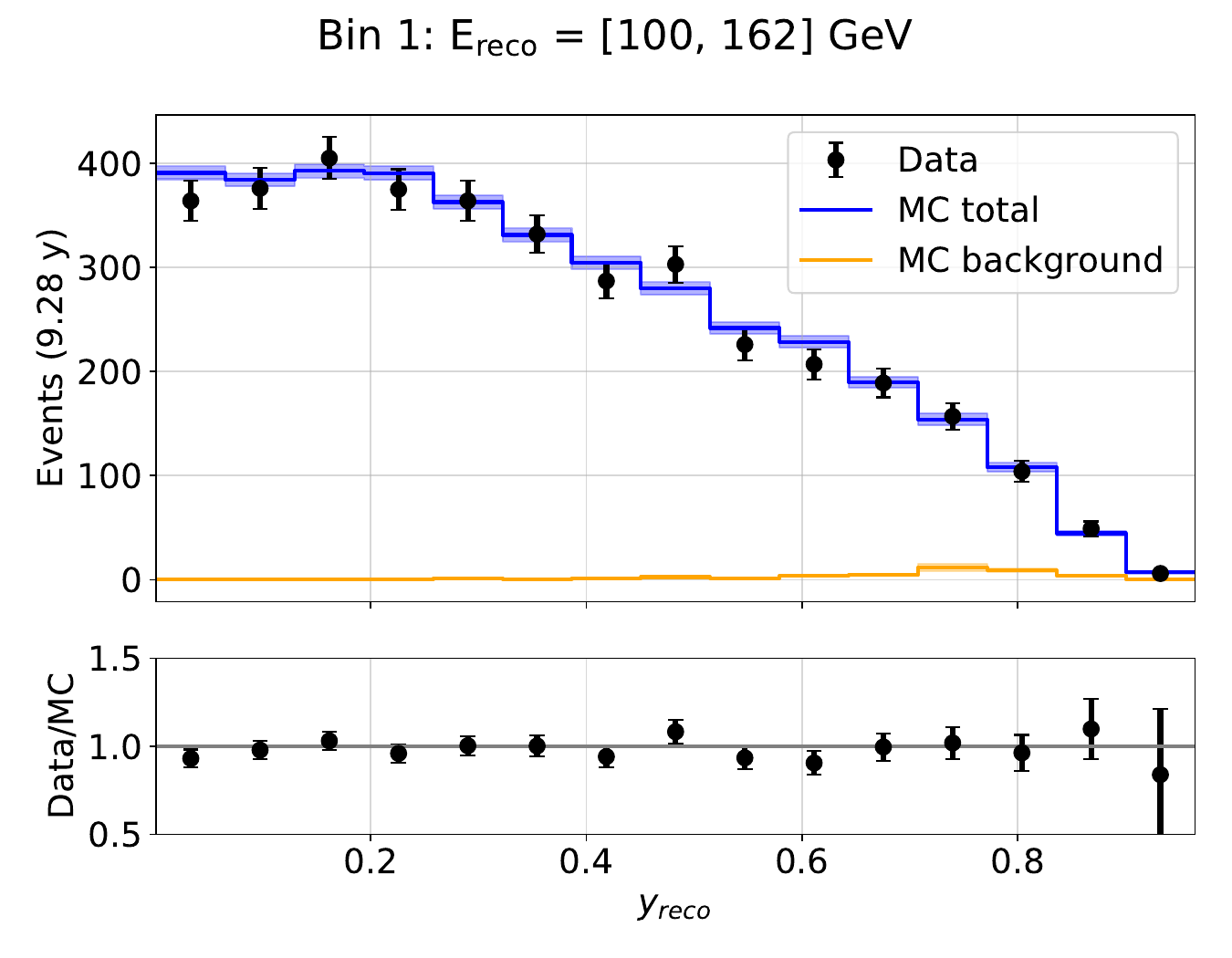}
    \includegraphics[width=0.3\linewidth]{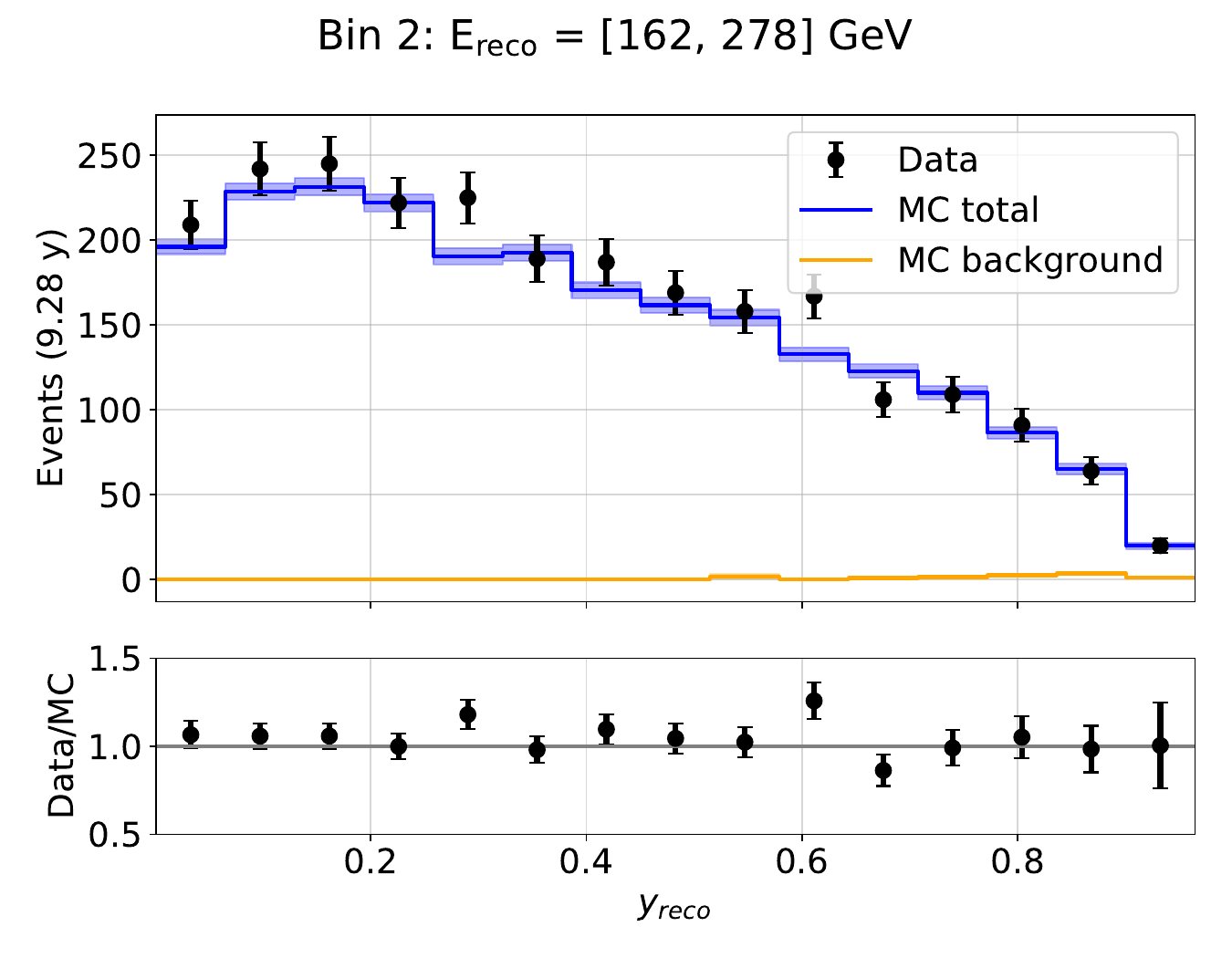}
    \includegraphics[width=0.3\linewidth]{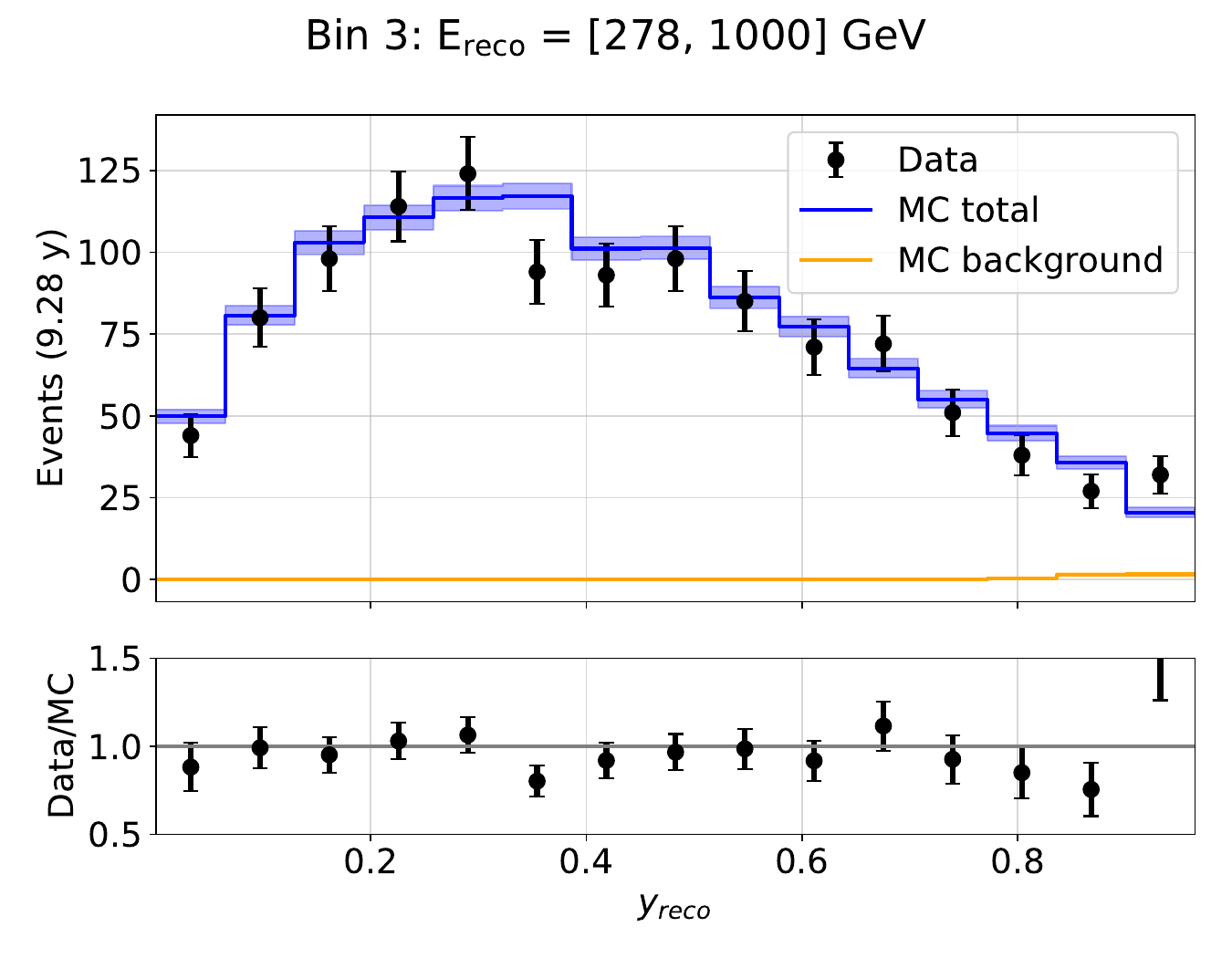}
    \caption{Data/Monte Carlo agreement post-fit for the reconstructed inelasticity distribution in each of three \Ereco bins.}
    \label{fig:data_mc_1d}
\end{figure*}

This analysis is performed by passing data and simulation through the same selection and reconstruction steps, and then comparing their distributions in the space of observables chosen. The shape of the true inelasticity distribution of the simulation is parametrized, so it can be modified, along with nuisance parameters that describe the impact of different sources of uncertainties, in order to find the values that result in the best fit to the data. In this study, we compare a two-dimensional histogram in \Ereco and \yreco with three bins in reconstructed energy between 100~GeV and 1~TeV and 11 equally sized bins in reconstructed inelasticity, between 0 and 1.

Figure~\ref{fig:energy_binning} shows the true energy distribution for each one of the three bins in reconstructed energy, assuming the {\sc genie} 2.12.8 cross section model~\cite{Andreopoulos:2015wxa} and HKKMS'15 atmospheric neutrino flux model by Honda~{\it et~al.}~\cite{Honda:2015fha}. The shaded bands indicate the central 68\% quantile, which is used to report the range in which the results are valid, while dashed lines indicate the median energy of the bin. Figure~\ref{fig:genie_template} shows the expectation from simulation for event counts in the analysis binning.

\subsection{Parametrization of inelasticity}
\label{sec:parameterization}
To parametrize the shape of the inelasticity distribution, our study takes advantage of the expected dependence of the neutrino-nucleon interaction cross section on inelasticity, as described in Sec.~\ref{sec:intro}. The functional form, used in the previous IceCube inelasticity study~\cite{IC_inelasticity_paper} and also described in Refs.~\cite{Binder_thesis:2017rlx,Klein_Spencer_book:2019nbu}, is the probability distribution of inelasticity and depends on two parameters $\epsilon$ and $\lambda$, 
\begin{equation}\label{eq::param_dpdy}
    \frac{dp}{dy}(\epsilon, \lambda) = \frac{1}{\sigma}\frac{d\sigma}{dy}(\epsilon, \lambda) =  N\cdot (1 + \epsilon (1-y)^2)\cdot y^{\lambda-1},
\end{equation}
where $\sigma$ is total cross section for a given energy, $\frac{d\sigma}{dy}$ is the differential cross section as function of inelasticity for a given energy, and $N$ is a normalization factor. 
To treat the distribution as a probability density function it is normalized to 1, so
\begin{equation}
    N = \frac{\lambda (\lambda+1)(\lambda+2)}{2\epsilon + (\lambda+1)(\lambda+2)}.
\end{equation}
Moreover, the parametrization can be transformed from $f(\epsilon, \lambda)$ into $f(\langle y\rangle, \lambda)$ via

\begin{equation}
    \langle y\rangle = \frac{\lambda (2\epsilon + (\lambda+2)(\lambda+3))}{(\lambda+3)(2\epsilon + (\lambda+1)(\lambda+2))},
\end{equation}

\noindent
where \meany is the mean of the inelasticity distribution. Since this quantity has a simple physical interpretation, the analysis is done in terms of \meany and $\lambda$.

It is important to emphasize that this parametrization is an effective description of the range of ``reasonable'' inelasticity distribution shapes; however, it has a number of limitations. To overcome these limitations, we use the \lglam in our analysis and in the following sections.
More details on this are provided in Appendix~\ref{apx:param_lim}.

The parametrization can be applied to both neutrino and antineutrino inelasticity distributions separately. 
Figure~\ref{fig::app::fit_to_step1_MC} shows Eq.~(\ref{eq::param_dpdy}) fitted to muon neutrino and antineutrino expectation from {\sc genie} for three energy bins. 
However, due to our inability to differentiate between neutrino and antineutrino events, we use the following parametrization to describe the flux-averaged inelasticity distribution at energy $E_\nu$,

\begin{equation}\label{eq::fl_av_dpdy}
\begin{split}
     \frac{dp}{dy}_{\mathrm{fl.~av.}}(E_\nu) =&~ \widetilde{\Phi}_\nu(\Phi_\nu,\sigma_\nu,\mathrm{sel.}; E_\nu) \cdot \frac{dp}{dy}_\nu(E_\nu) \\ &+ \widetilde{\Phi}_{\bar{\nu}}(\Phi_{\bar{\nu}},\sigma_{\bar{\nu}},\mathrm{sel.}; E_\nu) \cdot \frac{dp}{dy}_{\bar{\nu}}(E_\nu),
 \end{split}
\end{equation}
\noindent
where $\widetilde{\Phi}_{\nu(\bar{\nu})}$ is an energy-dependent fraction of neutrinos (antineutrinos) in our sample, which depends on atmospheric neutrino (antineutrino) flux $\Phi_{\nu(\bar{\nu})}$, the total neutrino-(antineutrino-)nucleon cross section $\sigma_{\nu(\bar{\nu})}$, and event selection, which is indicated as $\mathrm{sel.}$ in the equation. $\frac{dp}{dy}_{\nu(\bar{\nu})}$ denotes individual neutrino (antineutrino) inelasticity distributions. It is important to highlight that in Eq.~(\ref{eq::fl_av_dpdy}) $E_\nu$ refers to both neutrino and antineutrino energy.

\subsection{Fit of inelasticity distribution to data}

The analysis is performed using a \textit{forward folding} method~\cite{IceCubeCollaboration:2023wtb,IceCube_old_osc:2017lak} where the final level simulation is weighted at each combination of parameters tested and is subsequently used to populate the histogram that is compared to data. We employ an optimization algorithm to select the values of the parameters that will be tested and minimize the function
\begin{equation}
    \log\mathcal{L} = \sum_{i=1}^{N_{bins}} (k_i \log \mu_i - \mu_i) - \sum_{j=0}^{M_{priors}} \frac{(p_j - \hat{p_j})^2}{2 \sigma^2_j} 
\end{equation}
where the sum runs over all the bins of the histogram $N_{bins}$, $\mu_i$ stands for the number of expected events in the $i$th bin, which depends on the combination of parameters chosen, and $k_i$ is the number of observed events. The second term, where the sum is over the number of parameters with Gaussian (half-Gaussian) priors $M_{\mathrm{priors}}$, introduces prior knowledge on some of the fit parameters $p_j$ ($j=0,...,M_{\mathrm{priors}}$), penalizing deviations from their best estimates $\hat{p_j}$ in terms of their variance $\sigma^2_j$.

The neutrino simulation was produced using the {\sc genie} Monte Carlo generator~\cite{Andreopoulos:2015wxa} and weighted according to the expected atmospheric neutrino flux. Additional weighting factors were introduced for neutrinos and antineutrinos to produce flat \dpdy distributions, which could then be modified using Eq.~(\ref{eq::param_dpdy}) for all events in the same \Ereco bin. Reweighting all events to the same parametrized distribution means that we will be extracting flux-averaged \dpdy as shown in Eq.~(\ref{eq::fl_av_dpdy}).

We use the HKKMS'15 atmospheric neutrino flux model~\cite{Honda:2015fha} and include the effects of neutrino oscillations using the two-flavor approximation ($\nu_\mu \rightarrow \nu_\tau$) and the parameters reported in Ref.~\cite{Esteban:2018azc}, which is sufficient to describe the data since the low-energy threshold of this study is $E_{\mathrm{reco}}$=100~GeV. Atmospheric muons were produced using MuonGun, weighted to follow the expected flux from Refs.~\cite{muongun_two_Gaisser:2011klf,muongun_flux_Ahn:2009wx}.

\begin{table*}
\centering
\begin{tabular}{|l|l|l|l|l|l|}
\hline
\textbf{Group}                     & \textbf{Parameter} & \textbf{Best-fit value} & \textbf{Nominal value} & \textbf{Prior} \\ \hline
\multirow{6}{*}{\textbf{Physics parameters}} & $\langle y \rangle$ bin 1 & $0.45\pm0.02\pm0.005$ & 0.5 & Uniform  \\
                                   & $\langle y \rangle$ bin 2 & $0.46\pm0.02\pm0.004$ & 0.5 & Uniform    \\
                                   & $\langle y \rangle$ bin 3 & $0.45\pm0.02\pm0.006$ & 0.5 & Uniform    \\
                                   & $\log_{10}\lambda$ bin 1  & $0.02^{+0.09}_{-0.12}$ & 0.0 & Half-Gaussian (-0.4)   \\
                                   & $\log_{10}\lambda$ bin 2  & $0.10^{+0.06}_{-0.05}$ & 0.0 & Half-Gaussian (-0.4)   \\
                                   & $\log_{10}\lambda$ bin 3  & $0.10^{+0.05}_{-0.04}$ & 0.0 & Half-Gaussian (-0.4)  \\ \hline
\textbf{Flux systematics}                      & $\Delta \gamma_\nu$        & 0.11                    & 0.00                    & Gaussian ($\pm$ 0.1)         \\ \hline
\multirow{5}{*}{\textbf{Detector systematics}} & DOM efficiency            & 1.07                    & 1.00                    & Gaussian ($\pm$ 0.1)       \\
                                   & Ice absorption      & 1.00                    & 1.00                    & Uniform       \\
                                   & Ice scattering  & 1.01                    & 1.00                    & Uniform      \\
                                   & Relative efficiency $p_0$       & -0.5991                    & -0.2674                & Gaussian ($\pm$ 0.6)     \\
                                   & Relative efficiency $p_1$       & -0.02251                    & -0.04206               & Gaussian ($\pm$ 0.12)  \\ \hline
\textbf{Normalization}             & $A_{\mathrm{eff}}$ scale         & 0.78                    & 1.00                    & Uniform      \\ \hline
\end{tabular}
\caption{Summary of the best-fit values and settings for both the physics and nuisance parameters used in the analysis. All physics parameters are presented with $1\sigma$ uncertainties calculated using Wilks' theorem. The $\langle y \rangle$ parameters for each energy bin include an additional uncertainty due to limited MC statistics, as described in Sec.~\ref{sec:analysis:syst}.}
\label{tab:fit_results}
\end{table*}

The analysis was implemented in the \href{https://github.com/icecube/pisa}{{\sc pisa}}~\cite{pisa} framework. The optimization was done using the differential evolution global minimization method~\cite{Storn:1997uea} in order to be able to explore the parameter space thoroughly and also to introduce parameter-dependent boundaries, which are needed as described in Sec.~\ref{sec:parameterization} and Appendix~\ref{apx:param_lim}.

\subsection{Systematic uncertainties impacting the study}
\label{sec:analysis:syst}

The impacts of all the sources of systematic uncertainties that affect this analysis were introduced as correction factors that modify the weight of simulated events. By changing the weights, these factors modify the expectation of all the bins simultaneously, producing a distinct signature in the \Ereco vs \yreco space. These signatures differ from the ones produced by changes in the inelasticity distribution and therefore allow us to perform the measurement accurately, notwithstanding the uncertainties.

We considered all the sources of uncertainties studied in previous DeepCore analyses, using Ref.~\cite{IceCubeCollaboration:2023wtb} as our starting point. We studied their impact on the inelasticity parameters by estimating fit mismodeling and impact on the sensitivity using simulations and defined a subset of systematic uncertainties to be included in the fit to the data. The sources of uncertainties that remain are our limited understanding of the scattering and absorption lengths in the Antarctic ice, the effects of the refrozen ice columns on the angular acceptance of the DOMs, the absolute light collection efficiency of the modules, and the normalization and possible corrections to the energy dependence of the atmospheric neutrino flux. 
While a different set of inelasticity distribution parameters is fitted for each of the three \Ereco bins, the nuisance parameters are common for all bins.

The impact of ice and detector properties are obtained by simulating Monte Carlo sets with different input values and parametrizing the changes they produce in the bin counts of the analysis histogram. The five quantities used to parametrize said properties are scaling factors for DOM efficiency, ice absorption and scattering coefficients, and two parameters (relative efficiency $p_0$ and $p_1$), which were obtained using a principle component analysis of the DOM angular acceptance calibration data and are used to modify angular acceptance of the modules. The flux normalization is introduced as a global scaling factor, while the energy dependence is modeled as a power-law modification with a spectral index $E^{\Delta \gamma}$. The details of the implementation of these parameters can be found in Ref.~\cite{IceCubeCollaboration:2023wtb}.

Table~\ref{tab:fit_results} presents the parameters fitted in the study, including their nominal expectation and the prior knowledge, when applicable. The normalization is given a uniform prior, while the other systematic parameters that encode sources of uncertainty get a weak Gaussian prior. Additional information on the choice of priors is given in Ref.~\cite{IceCubeCollaboration:2023wtb}.

The limited Monte Carlo statistics introduce fluctuations in the selection efficiency for neutrinos and antineutrinos, which can potentially lead to a small bias in the fitted mean inelasticity. To account for this, we include an additional error term when reporting the \meany results for a given energy bin, such that
\begin{equation}
    \langle y \rangle = \langle y \rangle_{\mathrm{best~fit}} \pm \sigma_{\mathrm{Wilks'~th.}} \pm \Delta_{\mathrm{MC~stat.}},
\end{equation}
where $ \sigma_{\mathrm{Wilks'~th.}} $ is the $1\sigma$ interval obtained from the likelihood profile using the Wilks' theorem, and $ \Delta_{\mathrm{MC~stat.}} $ is the bias in \meany obtained using results of an inject-recover test to a sample where original {\sc genie} inelasticity distribution was not modified by reweighting. The sample was fitted using the parametrization in Eq.~(\ref{eq::param_dpdy}), and the best-fit \meany value for each bin was compared to the expectation for the same energy bin, calculated using generator-level simulation, with the difference being assigned to the $ \Delta_{\mathrm{MC~stat.}} $ term. Values of the $ \Delta_{\mathrm{MC~stat.}} $ for each of the energy bins is shown in Table~\ref{tab:fit_results}.

\section{Results~\label{sec:results}}
\begin{figure*}
    \centering
    \includegraphics[width=0.31\linewidth]{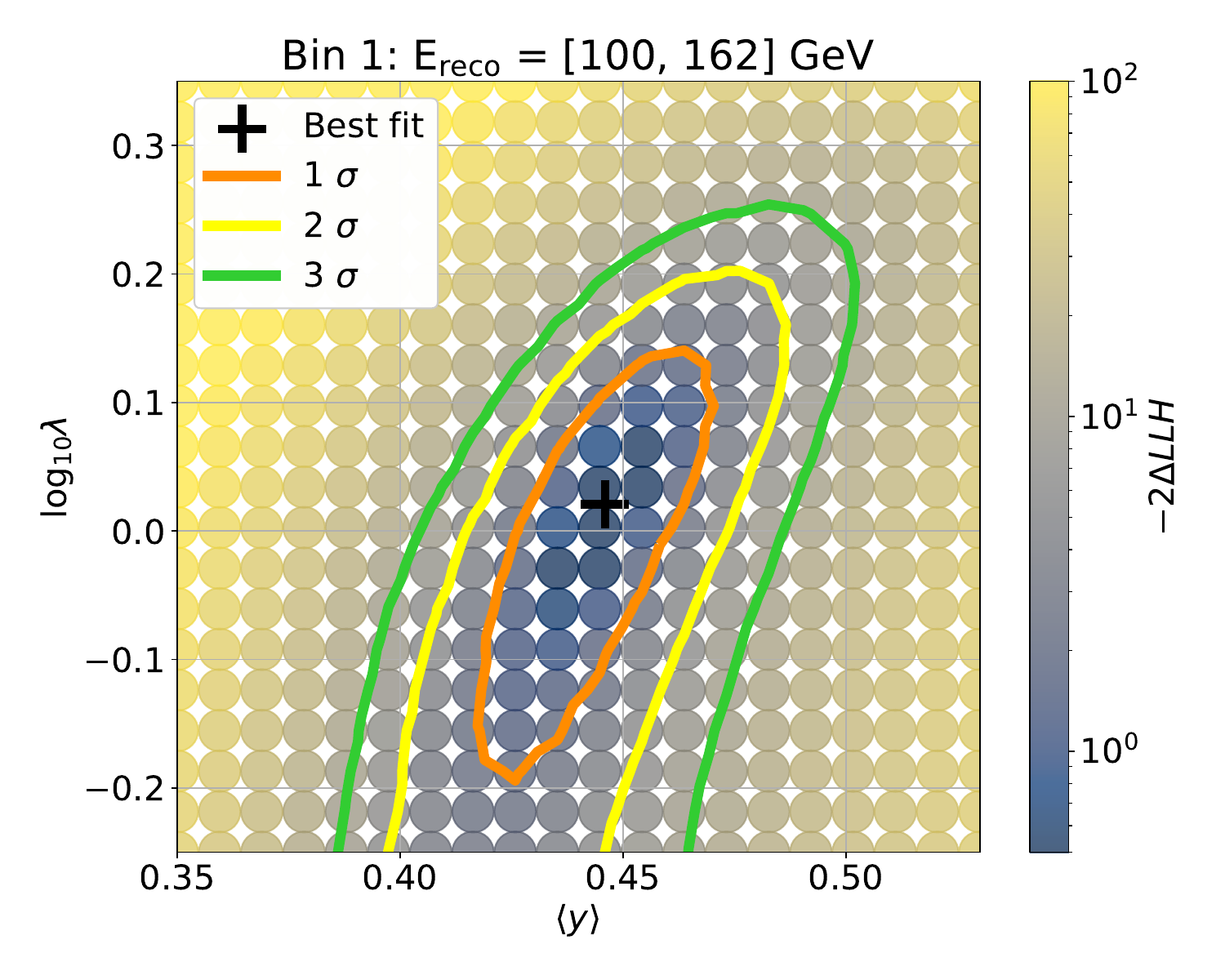}
    \includegraphics[width=0.31\linewidth]{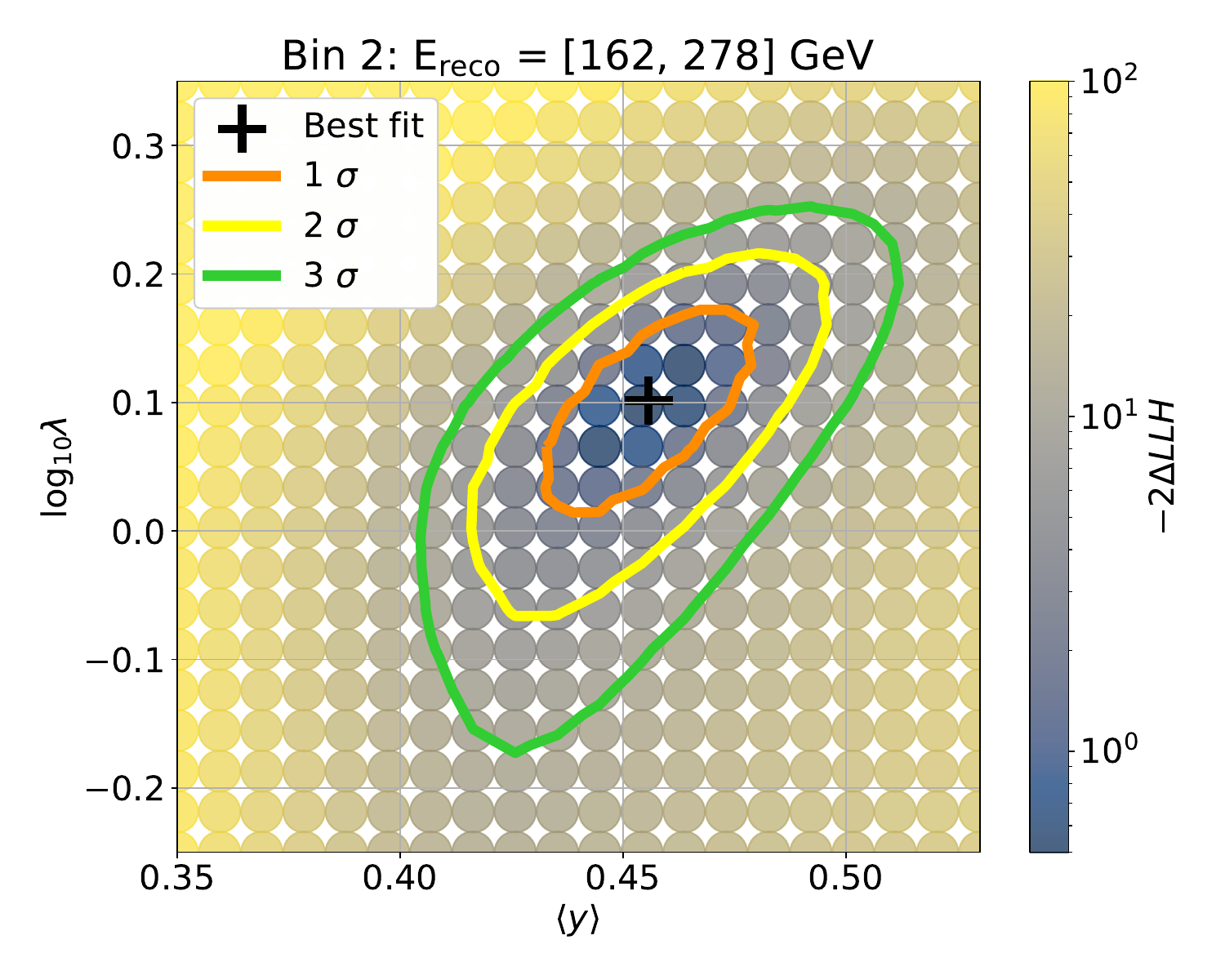}
    \includegraphics[width=0.31\linewidth]{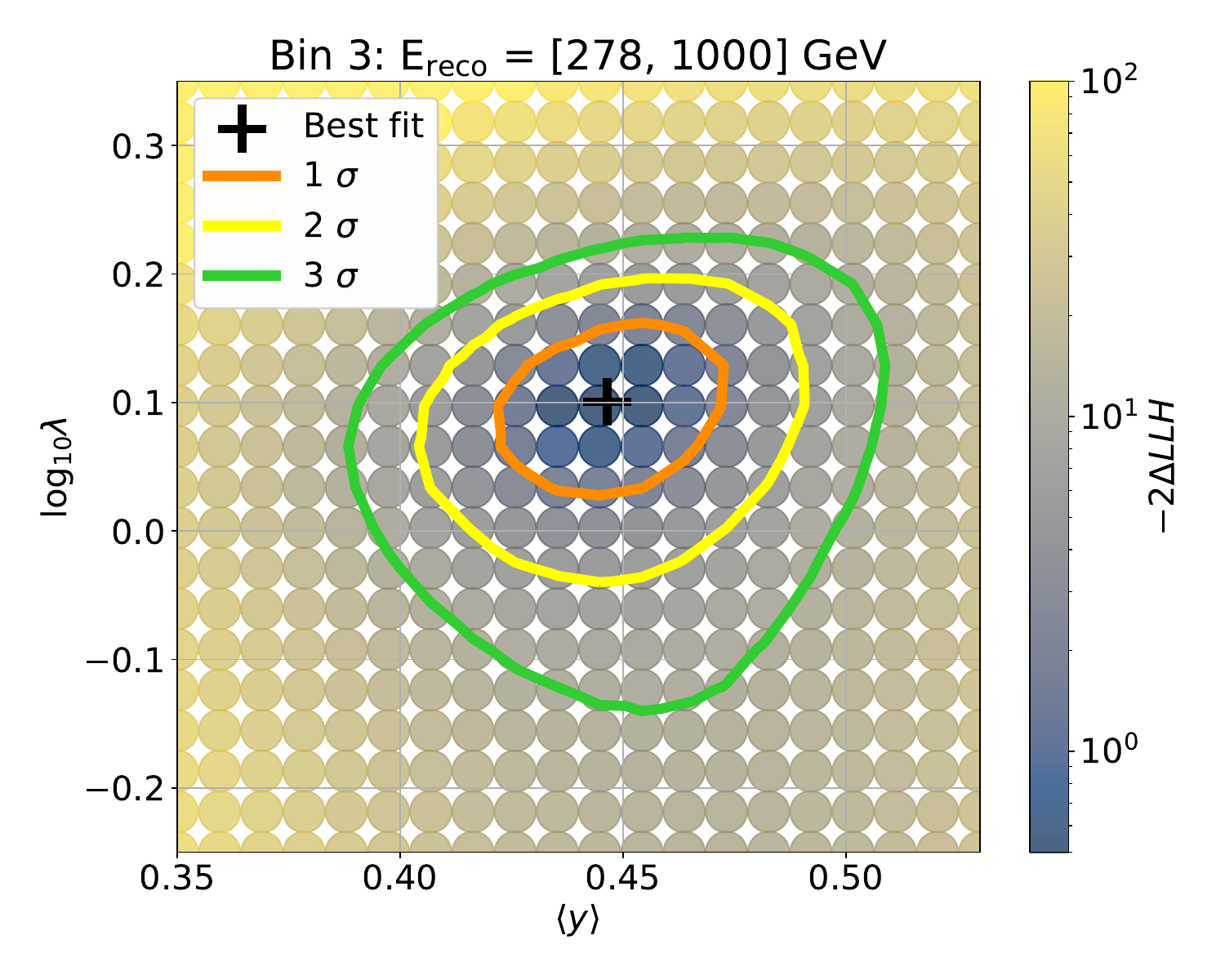}
    \includegraphics[width=0.31\linewidth]{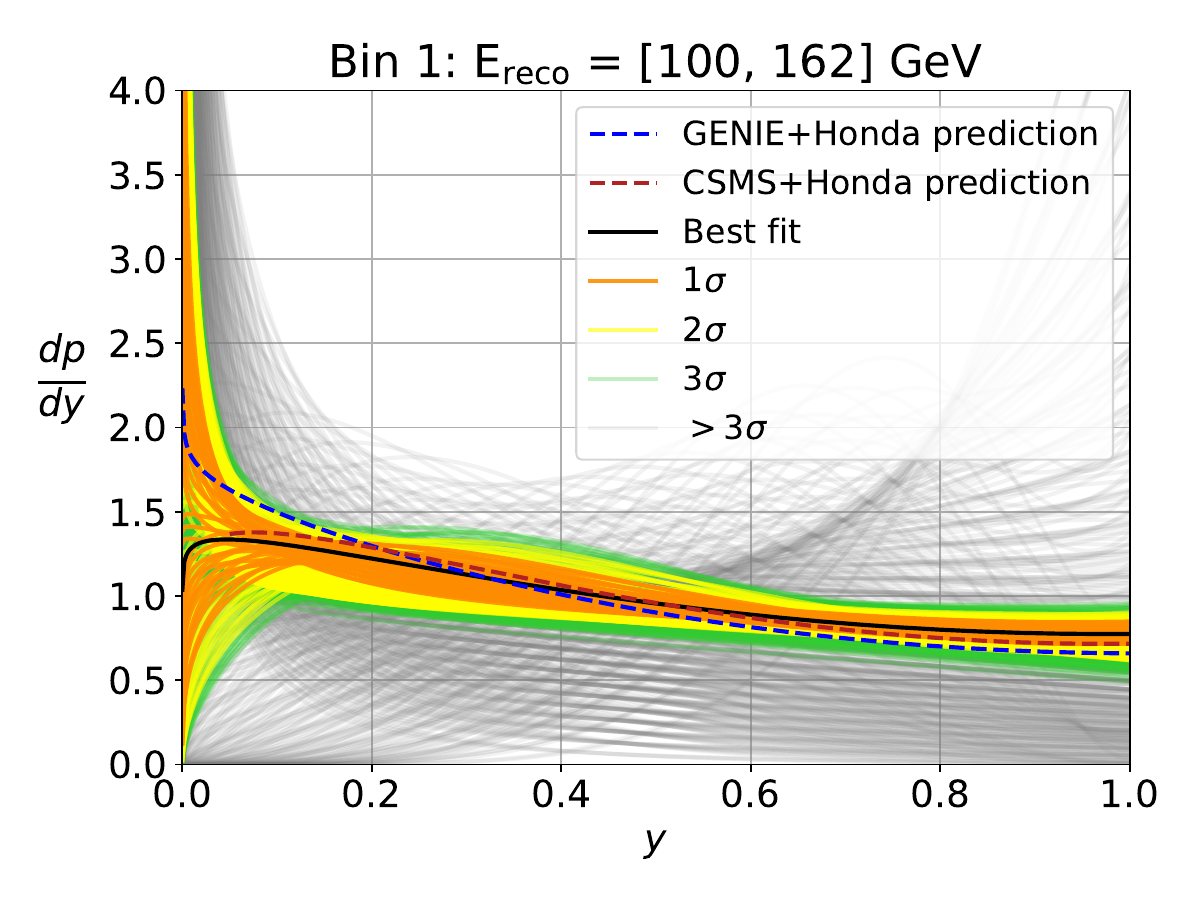}
    \includegraphics[width=0.31\linewidth]{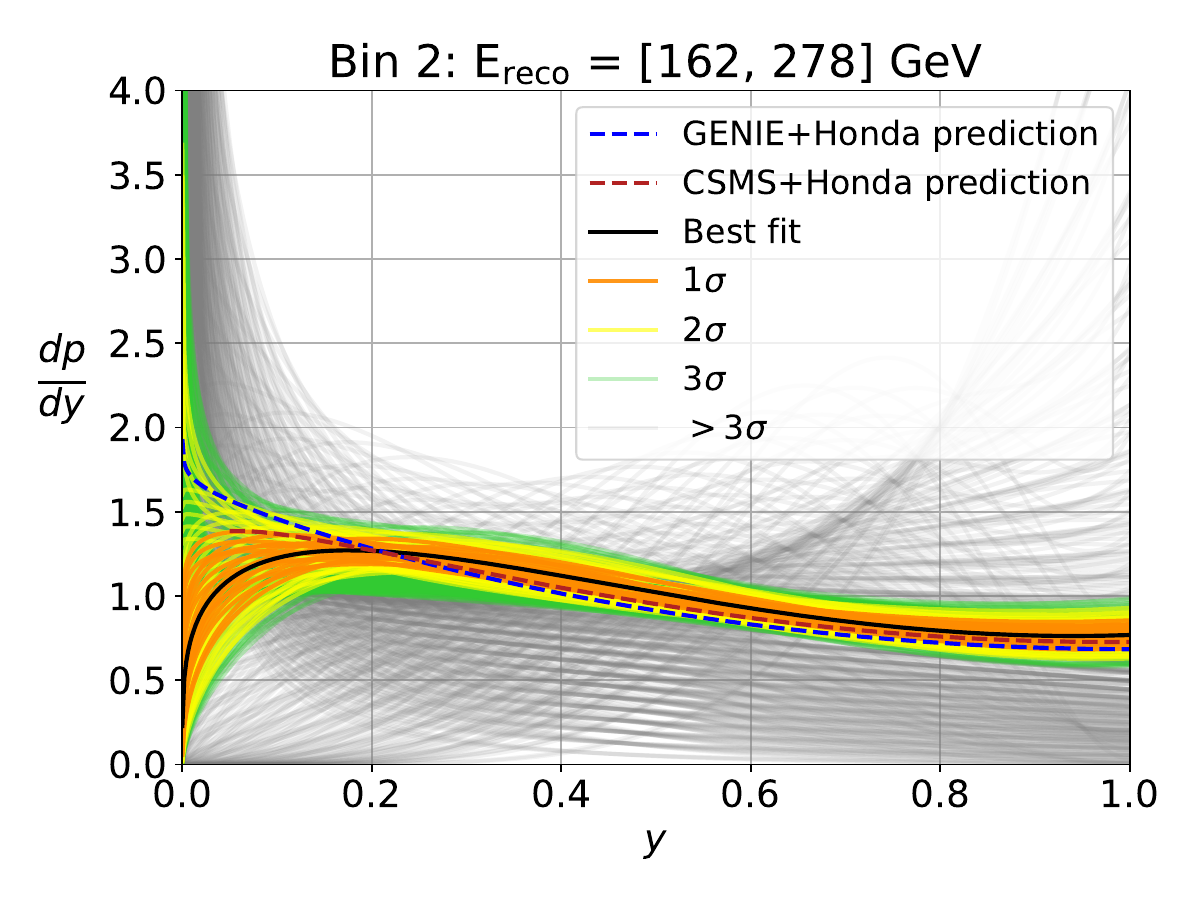}
    \includegraphics[width=0.31\linewidth]{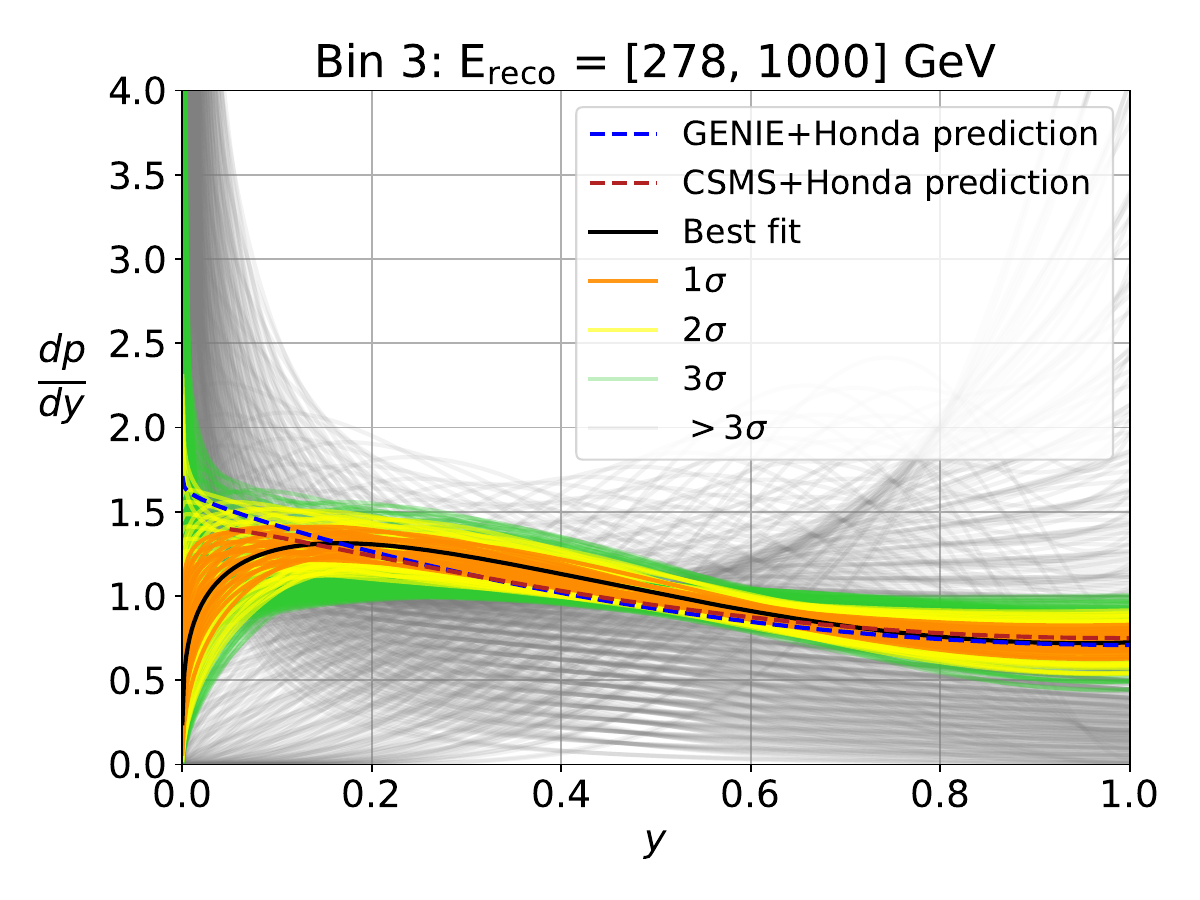}
    \caption{Top: likelihood space of the physical parameters $(\langle y \rangle, \log_{10}\lambda)$ with contours that denote the uncertainty at 1, 2, or 3~$\sigma$. Bottom: inelasticity distribution shapes as function of inelasticity that results from the fit, also showing the curves that correspond to the points within each confidence region displayed on the top panel in corresponding colors. The gray lines in the background show a representative sample of the values allowed during fit that are beyond 3~$\sigma$. The figure also includes predictions that result from the combination of {\sc genie} or CSMS cross section (both calculated for water) and the HKKMS'15 atmospheric neutrino flux calculation. }
    \label{fig:res_2d_scans}
\end{figure*}

We applied the analysis to data from the 2011--2021 seasons, which correspond to 9.28 years of detector livetime. The post-fit Monte Carlo is in good agreement with the data, with a $p$-value of 9.5\%. Figure~\ref{fig:data_mc_1d} shows this for \yreco in each of the three \Ereco bins. The mean inelasticity is compatible with a constant value \meany=~0.45 across all energies, with a $\lambda$ parameter that differs for the lowest energies. The values for all the fit parameters are shown in Table~\ref{tab:fit_results}. The top panel of Fig.~\ref{fig:res_2d_scans} shows two-dimensional confidence regions for \meany and \lglam in each \Ereco bin. The bottom panel on Fig.~\ref{fig:res_2d_scans} illustrates \dpdy distributions corresponding to the best fit as well as sampled points within each confidence region. We also show the expectation for our sample using both {\sc genie}~\cite{Andreopoulos:2015wxa} and Cooper-Sarkar--Mertsch--Sarkar (CSMS)~\cite{DIS:CSMS} cross section models, with the HKKMS'15 atmospheric neutrino flux model~\cite{Honda:2015fha}. The inelasticity distributions show slight differences in shape, with our fit preferring a relatively lower contribution at low~$y$ and a relatively higher contribution at medium and high~$y$ values.

\begin{figure}
    \centering
    \includegraphics[width=0.99\linewidth]{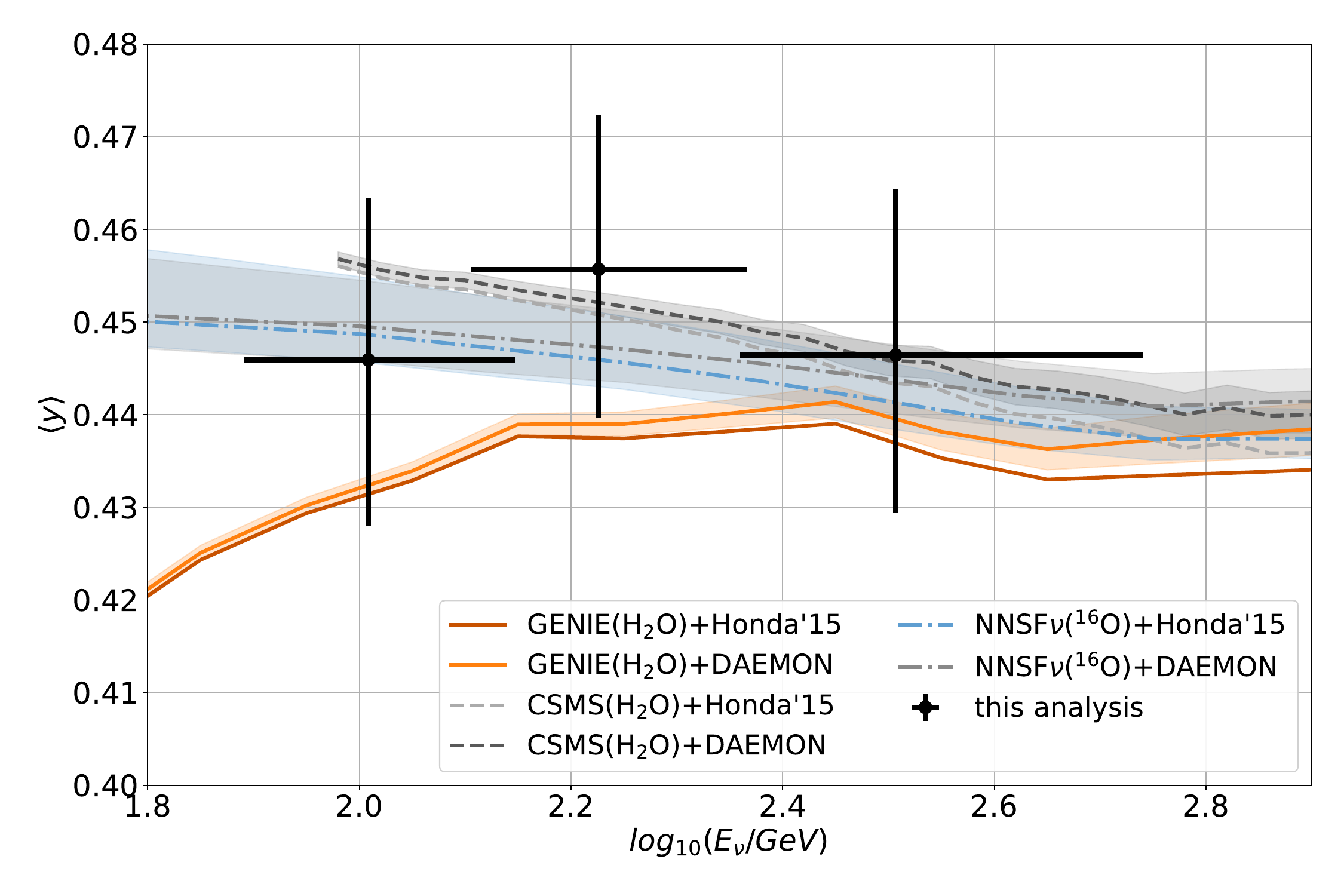}
    \caption{Comparison of measured mean inelasticity to predictions calculated using six different combinations of available cross section models~\cite{Andreopoulos:2015wxa,DIS:CSMS,NNSF_nu} and atmospheric neutrino flux models~\cite{Honda:2015fha,daemonflux}. The uncertainties were calculated using reported model uncertainties (available for NNFS$\nu$~\cite{NNSF_nu} cross section and {\sc daemonflux}~\cite{daemonflux}). }
    \label{fig:res_mean_y_models}
\end{figure}

\begin{figure}
    \centering
    \includegraphics[width=0.99\linewidth]{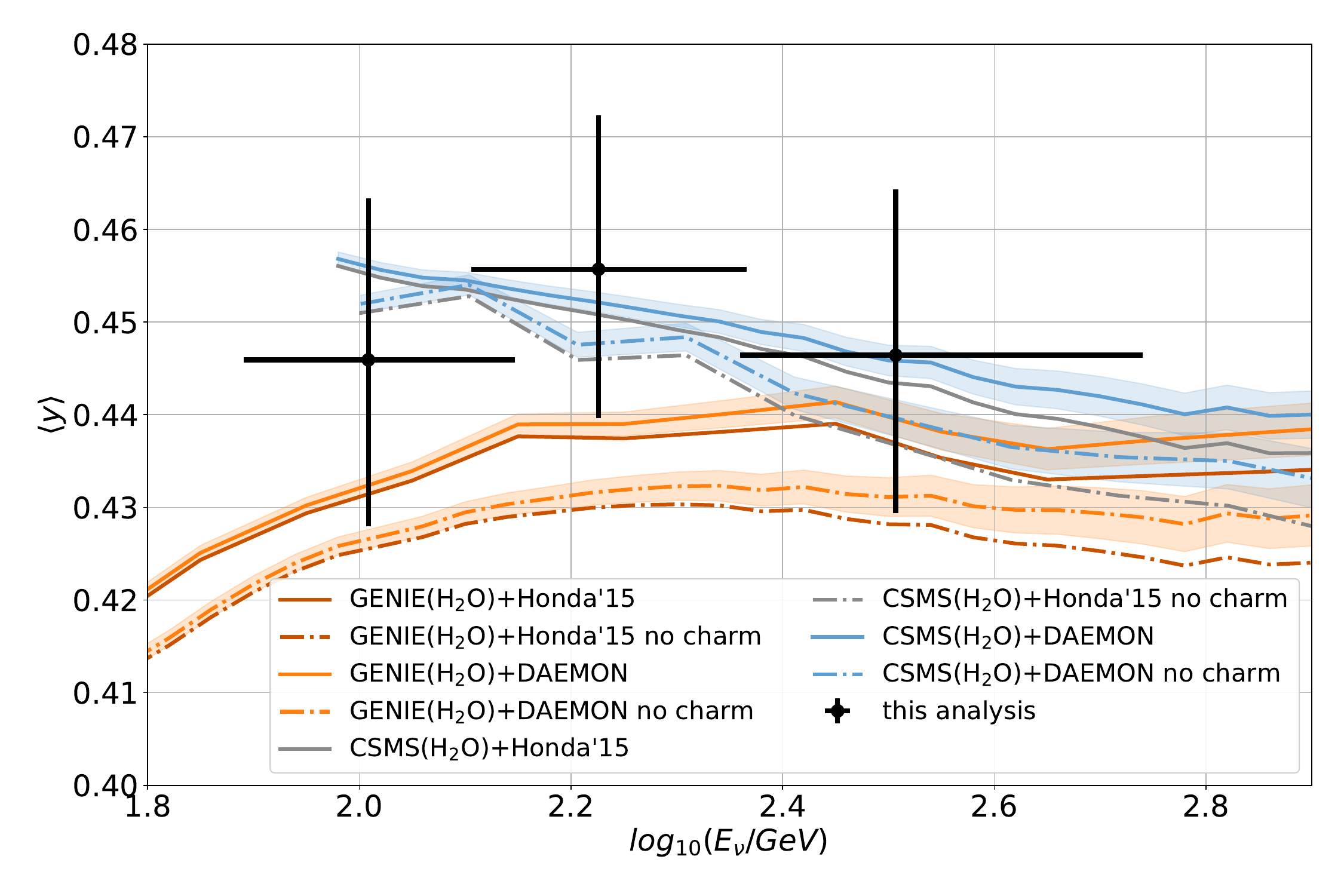}
    \caption{Comparison of measured mean inelasticity to predictions calculated using {\sc genie} and CSMS cross sections for the case of no charm production as well as for the nominal case; the predictions are calculated for both HKKMS and {\sc daemonflux} atmospheric neutrino flux models.}
    \label{fig:res_mean_y_models_charm}
\end{figure}

The mean inelasticity \meany as a function of energy is shown in Fig.~\ref{fig:res_mean_y_models}, compared to the expectation from various cross section calculations, namely, {\sc genie}~\cite{Andreopoulos:2015wxa}, CSMS~\cite{DIS:CSMS} and NNSF$\nu$~\cite{NNSF_nu}, and two atmospheric neutrino fluxes: HKKMS'15~\cite{Honda:2015fha} and {\sc daemonflux}~\cite{daemonflux}. It is important to mention that, aside from differences in the calculational approach and the parton distribution functions used, the different cross section models in this comparison make different assumptions about the target. The {\sc genie} cross section is calculated for ice (H2O) and takes into account nuclear shadowing and antishadowing effects~\cite{Andreopoulos:2015wxa}. The CSMS cross section is also calculated for an H2O target. For the NNSF$\nu$ cross section, we are using the prediction for an oxygen target. Another distinction is that {\sc genie} and NNSF$\nu$ are inclusive cross section models, while CSMS describes only the deep-inelastic scattering of neutrinos.
Recent works on the neutrino cross section~\cite{SPENCER_iso_paper,Weigel_iso_paper:2024gzh} show that the assumption of the isoscalar target (i.e., consisting of equal number of protons and neutrons) affects the expected inelasticity distribution for DIS events compared to that of water. We estimate that size of the isoscalar assumption effect on our sample will be smaller than precision of our results. We have also investigated the effect of not including nuclear shadowing and antishadowing corrections provided by {\sc genie} and found it to be much smaller than our measurement precision.
To highlight the impact of neutrino-induced charm production on the mean inelasticity prediction, we show in Fig.~\ref{fig:res_mean_y_models_charm} a comparison of the measured mean inelasticity to the {\sc genie} and CSMS models with and without neutrino-induced charm production in combination with both the HKKMS'15 and {\sc daemonflux}.

\begin{figure*}
    \centering
    \includegraphics[width=0.80\linewidth]{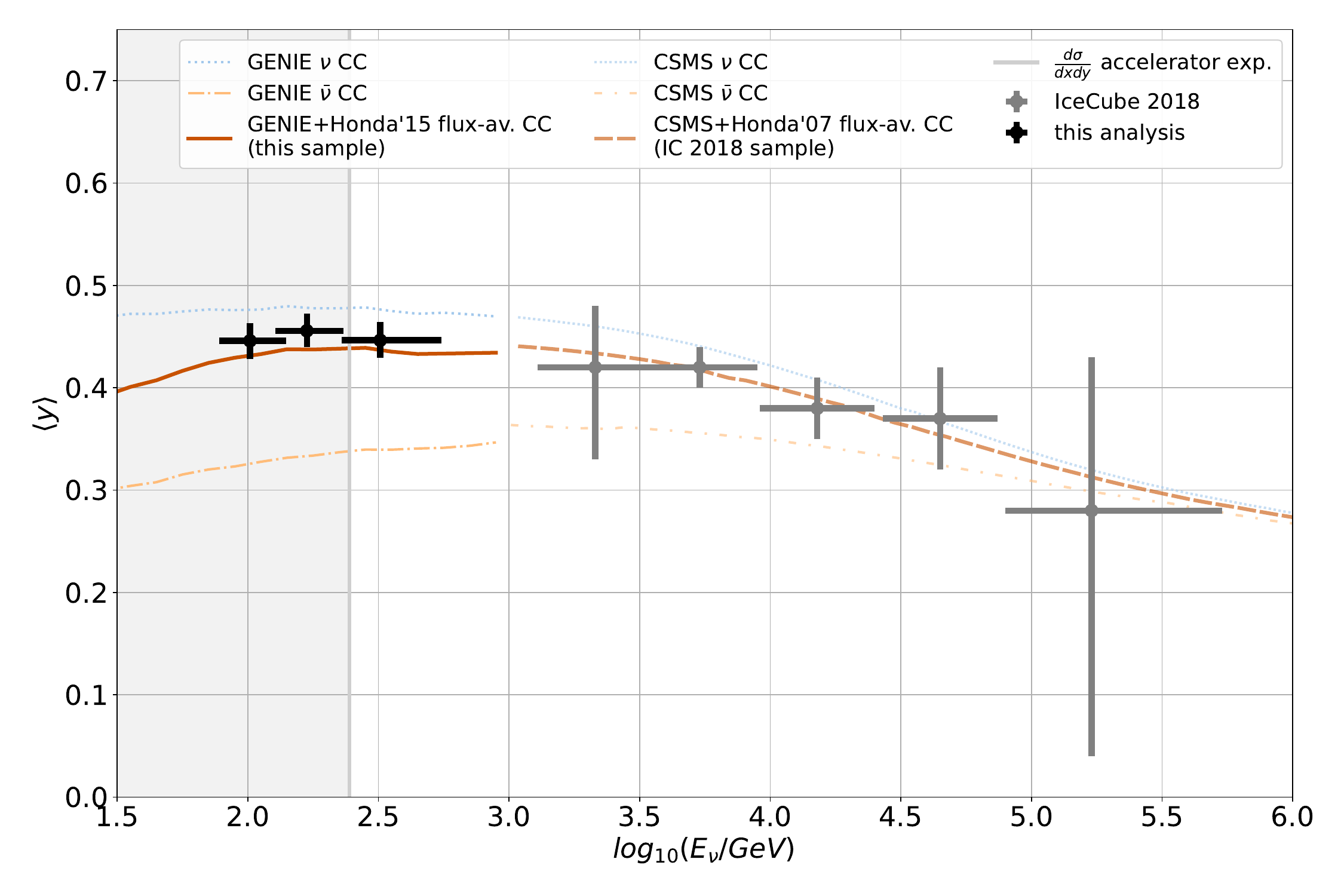}
    \caption{Measured mean inelasticity $\langle y \rangle$ as a function of neutrino energy is compared to the {\sc genie}~2.12.8~$+$~HKKMS~2015 prediction for our event sample. The IceCube 1~TeV--1~PeV inelasticity result~\cite{IC_inelasticity_paper} and the CSMS~\cite{DIS:CSMS}~$+$~HKKMS~2015 prediction for the corresponding sample are also shown. The highlighted energy range below 245~GeV is where measurements of differential cross section have been made by NuTeV/CCFR~\cite{NuTeV_diff_xsec} and CDHSW~\cite{CERN_diff_xsec}.}
    \label{fig:res_mean_y_large}
\end{figure*}

Figure~\ref{fig:res_mean_y_large} shows the mean inelasticity compared with results obtained using higher-energy interactions in IceCube~\cite{IC_inelasticity_paper}. Only a subset of the models tested are shown. It is important to note that the mean inelasticity prediction is dependent on the event selection, so the results are not directly comparable in detail, but the general trend is consistent.

\begin{figure}
    \centering
    \includegraphics[width=0.99\linewidth]{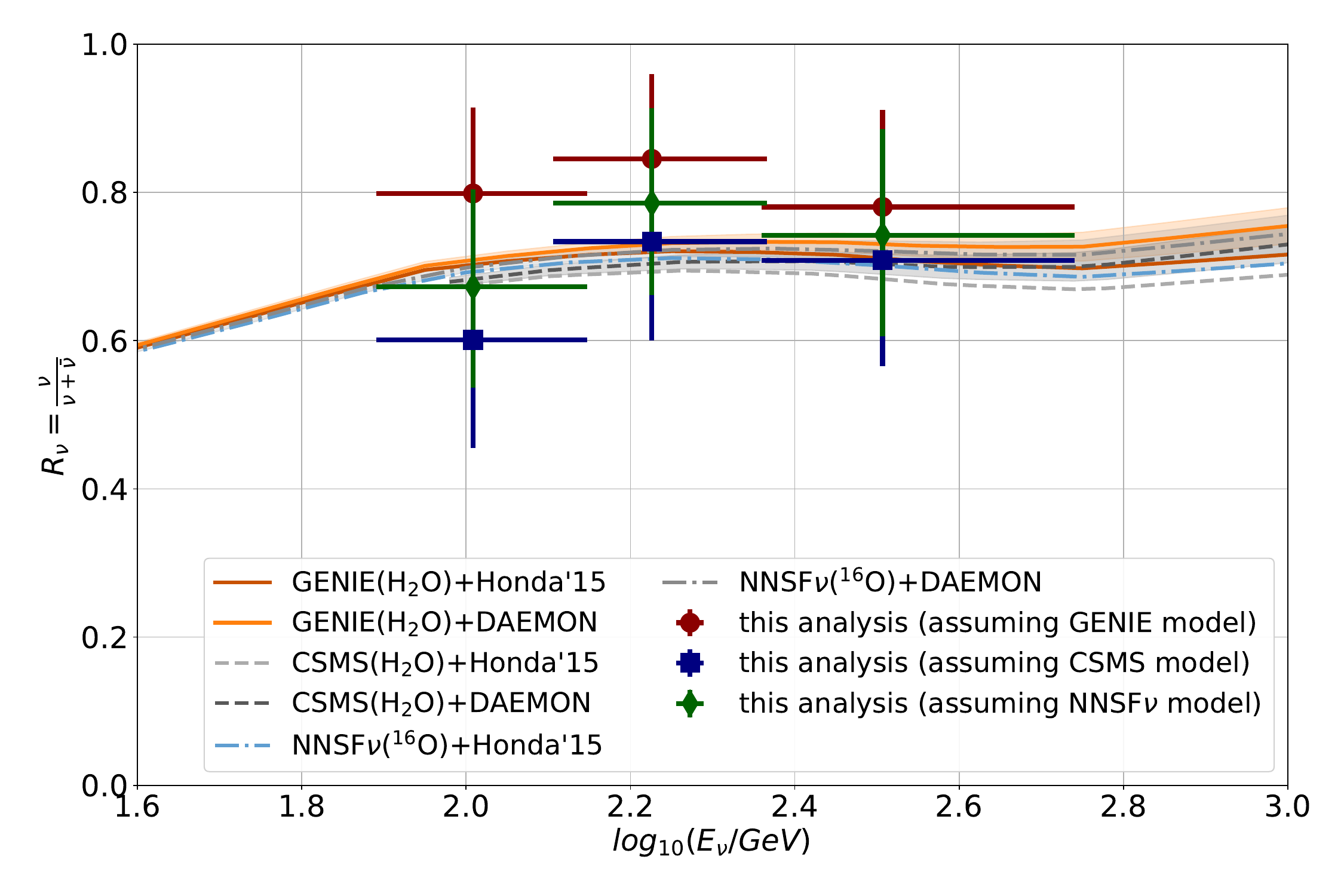}
    \caption{Neutrino fraction in the atmospheric neutrino flux calculated for our sample assuming different cross section models and compared to theoretical predictions.}
    \label{fig:res_nu_nubar}
\end{figure}

The mean inelasticity can also be interpreted as a measure of the ratio of neutrinos to antineutrinos in the atmospheric neutrino flux, if we assume a specific cross section. This comparison is shown in Fig.~\ref{fig:res_nu_nubar}, together with the neutrino-antineutrino fraction expected from the same combinations of models discussed before. The spread of the predictions is significantly smaller than the error bars of our result. 

\section{Conclusions}
\label{sec:discussion}

In this study, we used the observed energy in tracklike and cascadelike byproducts of neutrino candidate events in IceCube to extract the inelasticity of $\nu$-nucleon interactions. We focus on a highly pure sample of $\nu_\mu$-CC events between 80 and 560~GeV that interact inside the DeepCore volume. 
This is the first time that such details of the differential neutrino cross section have been explored at energies between 245~GeV and 1~TeV.

The cross section as a function of inelasticity was parametrized as before~\cite{IC_inelasticity_paper}, and we measured two parameters that define its shape \meany and $\lambda$. We then compared the mean inelasticity \meany obtained with the expectation from three cross section models and two atmospheric neutrino fluxes. 

The results of this study do not exclude any of the models at the 1$\sigma$ level. The study will improve with the addition of more years of detector data, and ongoing developments on event reconstruction that could improve the separation of tracks and cascades, as well as their energy estimation. 
The IceCube Upgrade~\cite{Ishihara_upgrade:2019aao} is a detector extension that will be deployed in the 2025--2026 season. It includes both new optical sensors and additional calibration devices which will be placed in the center of DeepCore. Future iterations of this study will benefit from tighter optical module spacing, which is expected to both improve reconstruction at the energy range of this analysis, as well as allow to extend it to neutrino energies below 80~GeV. In addition to this, new calibration devices will help us restrict the size of the systematic uncertainties on the detector response.

\begin{acknowledgements}
We thank Yuri Onishchuk for contributions to the early phase of this work.
The authors gratefully acknowledge the support from the following agencies and institutions:
in the USA, the U.S. National Science Foundation-Office of Polar Programs,
U.S. National Science Foundation-Physics Division,
U.S. National Science Foundation-EPSCoR,
U.S. National Science Foundation-Office of Advanced Cyberinfrastructure,
Wisconsin Alumni Research Foundation,
Center for High Throughput Computing (CHTC) at the University of Wisconsin{\textendash}Madison,
Open Science Grid (OSG),
Partnership to Advance Throughput Computing (PATh),
Advanced Cyberinfrastructure Coordination Ecosystem: Services {\&} Support (ACCESS),
Frontera and Ranch computing project at the Texas Advanced Computing Center,
U.S. Department of Energy-National Energy Research Scientific Computing Center,
Particle astrophysics research computing center at the University of Maryland,
Institute for Cyber-Enabled Research at Michigan State University,
Astroparticle physics computational facility at Marquette University,
NVIDIA Corporation,
and Google Cloud Platform;
in Belgium, Funds for Scientific Research (FRS-FNRS and FWO),
FWO Odysseus and Big Science programmes,
and Belgian Federal Science Policy Office (Belspo);
in Germany, Bundesministerium f{\"u}r Bildung und Forschung (BMBF),
Deutsche Forschungsgemeinschaft (DFG),
Helmholtz Alliance for Astroparticle Physics (HAP),
Initiative and Networking Fund of the Helmholtz Association,
Deutsches Elektronen Synchrotron (DESY),
and High Performance Computing cluster of the RWTH Aachen;
in Sweden, Swedish Research Council,
Swedish Polar Research Secretariat,
Swedish National Infrastructure for Computing (SNIC),
and Knut and Alice Wallenberg Foundation;
in the European Union, EGI Advanced Computing for research;
in Australia, Australian Research Council;
in Canada, Natural Sciences and Engineering Research Council of Canada, Calcul Québec, Compute Ontario, Canada Foundation for Innovation, WestGrid, Digital Research Alliance of Canada, and Arthur B. McDonald Canadian Astroparticle Physics Research Institute;
in Denmark, Villum Fonden, Carlsberg Foundation, and European Commission;
in New Zealand, Marsden Fund;
in Japan, Japan Society for Promotion of Science (JSPS)
and Institute for Global Prominent Research (IGPR) of Chiba University;
in Korea, National Research Foundation of Korea (NRF);
in Switzerland, Swiss National Science Foundation (SNSF).
\end{acknowledgements}

\appendix
\section{Limitations of the parametrization~\label{apx:param_lim}}
\begin{figure}
    \centering
    \includegraphics[width=0.99\linewidth]{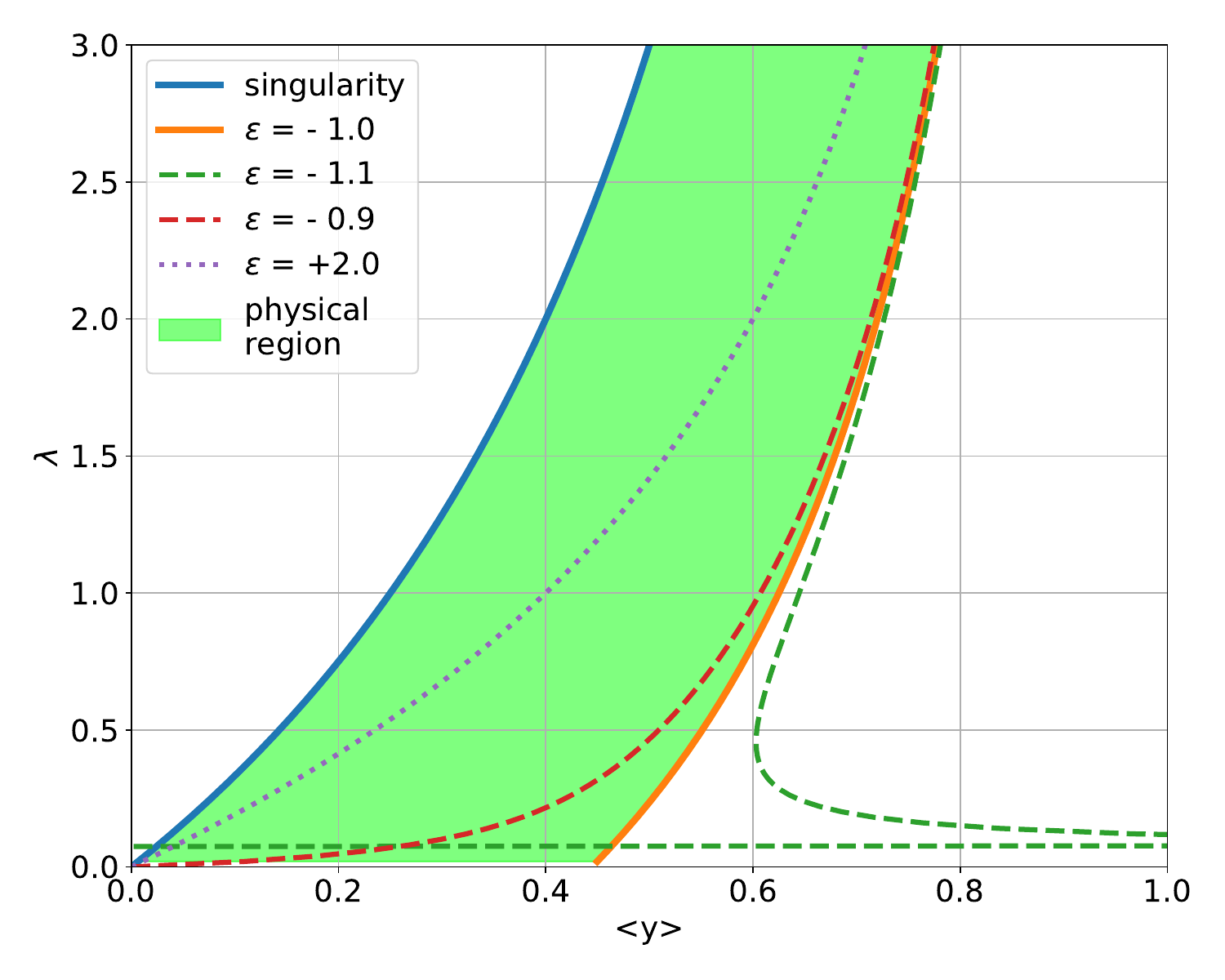}
    
    \caption{Physically allowed and forbidden regions in the physics parameter space.}
    \label{fig::unphys}
\end{figure}

The physical interpretation of Eq.~(\ref{eq::param_dpdy}) requires $y=[0,1]$, which means that $ 0 < $\meany$ < 1$, and $\frac{dp}{dy}$ must be positive, which in turn requires that $\lambda > 0$ and $\epsilon > -1$. The allowed physical space and these limits are shown in Fig.~\ref{fig::unphys}. 

A special case arises when $ \lambda < 1 $, because in that case $ \lim_{y\to 0} \frac{dp}{dy} = \infty $. Because of the finite number of events in our data, and also because of presence of kinematic constraints on events with very low inelasticity, the fit might not be able to detect a sharp rise in the distribution near $y=0$. Since there are no data in this region, the integral of the distribution diverges from $1$, thus violating the probability distribution function normalization requirement. This has two undesirable consequences in our analysis. First, this means that the \meany variable in Eq.~(\ref{eq::param_dpdy}) loses its physical interpretation as the mean inelasticity. The second outcome is that breaking of the normalization requirement can cause the minimizer to use \dpdy parameters to modify normalization of the data, which will absorb part of the systematic uncertainties in the study, affecting the correctness of the measurement.

To avert this problem, we adjusted the simulation to data in $\log_{10}\lambda$ instead of $\lambda$ and put a weak half-Gaussian prior on negative $\log_{10}\lambda$  with $\sigma(\log_{10}\lambda) = 0.4$ on the negative side. 
Then, if the data prefer $ \lambda < 1 $, the fit can still allow it with minimal bias, but if this is not the case, the prior will prevent the minimizer from getting stuck in local minima, which is especially relevant when performing a scan to determine confidence regions. The correct behavior of the fit in these cases was confirmed with Monte Carlo tests.

\section{Parameter impact on analysis sample~\label{apx:param_impact}}

\begin{figure}[t!]
    \centering
    \includegraphics[width=0.95\linewidth]{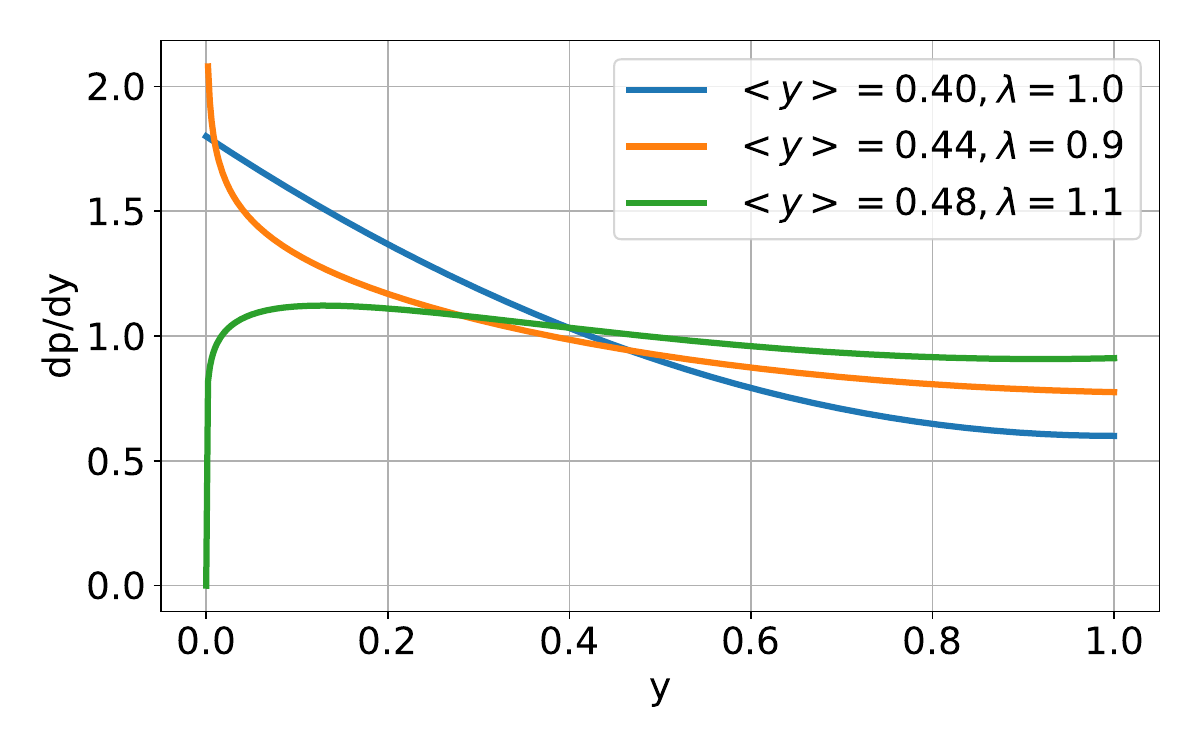}
    \caption{Example $\frac{dp}{dy}$ distributions for different combinations of parameters $\langle y\rangle$ and $\lambda$.}
    \label{fig:example_dpdy}
\end{figure}

\begin{figure*}[!t]
    \centering
    \includegraphics[width=0.31\linewidth]{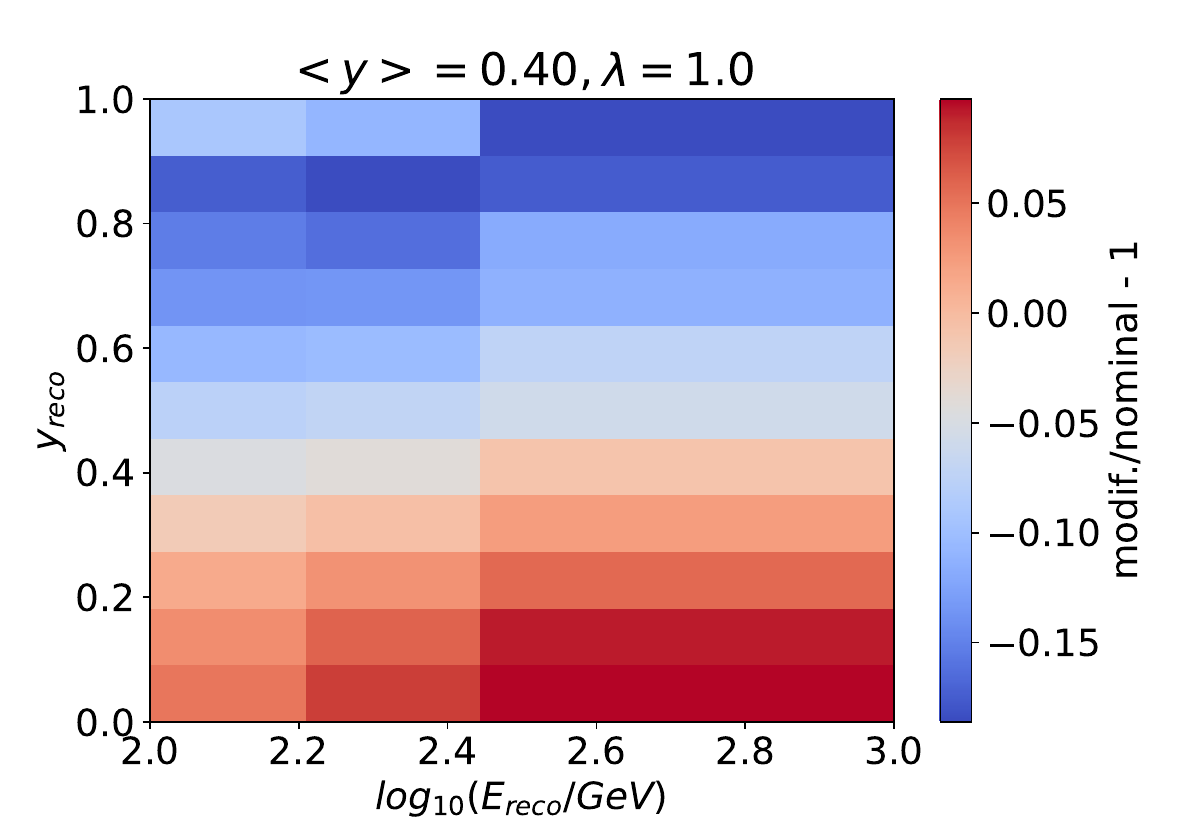}
    \includegraphics[width=0.31\linewidth]{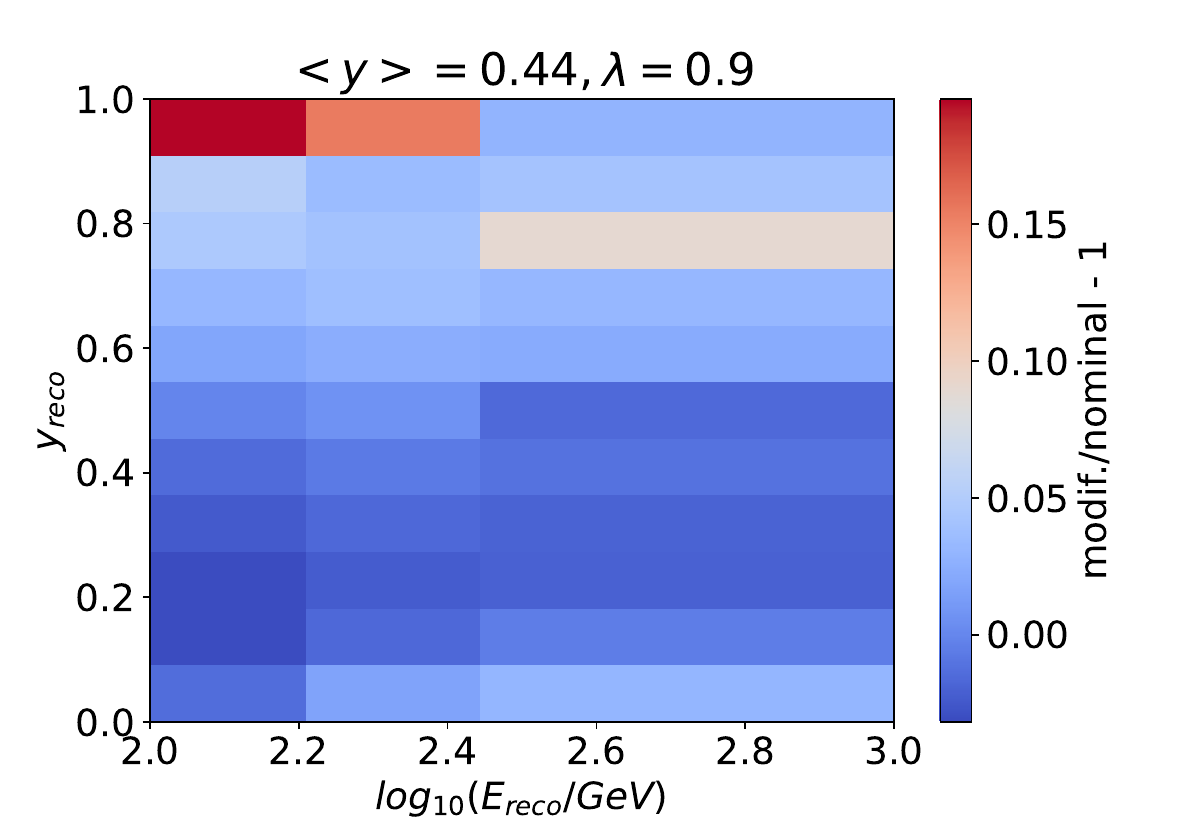}
    \includegraphics[width=0.31\linewidth]{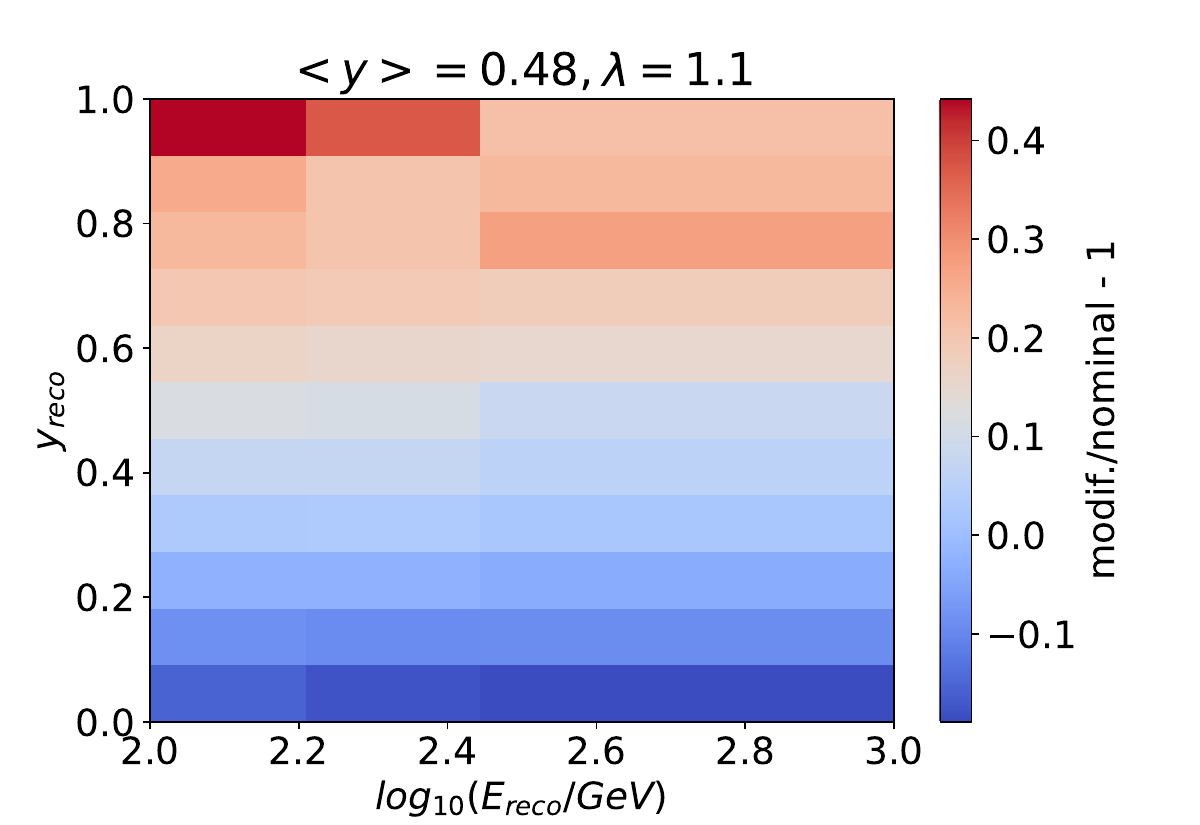}
    \caption{Modifications to the nominal expectation introduced when physics parameters in all energy bins are varied.}
    \label{fig:app:maps_modif}
\end{figure*}

\begin{figure*}
    \centering
        \includegraphics[width=0.31\linewidth]{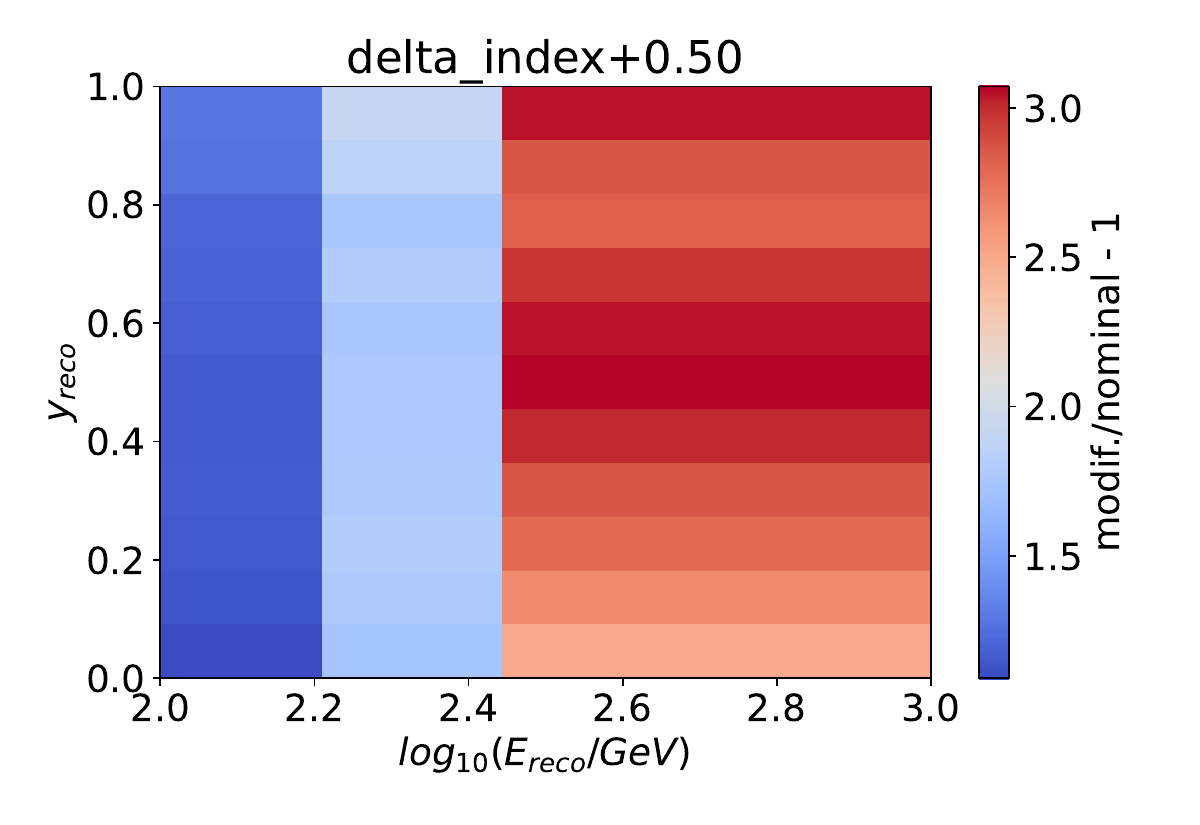}
        \includegraphics[width=0.31\linewidth]{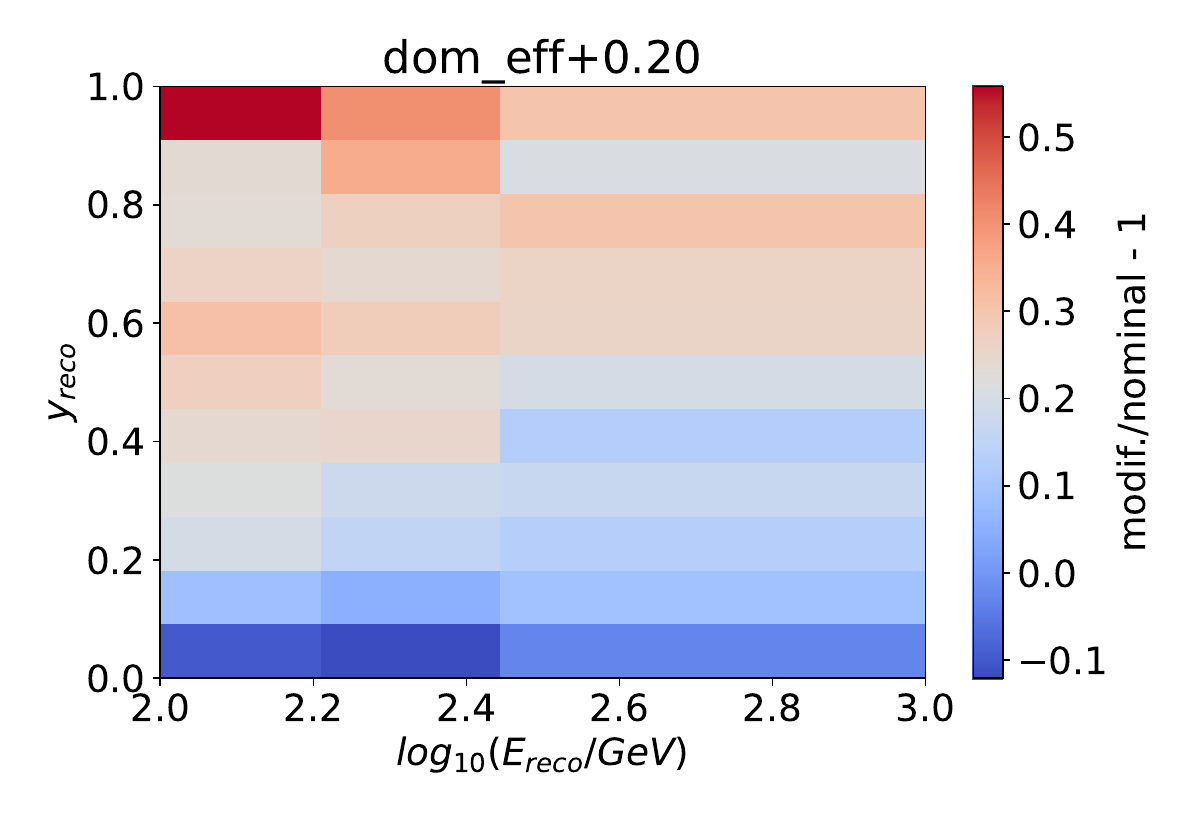}
        \includegraphics[width=0.31\linewidth]{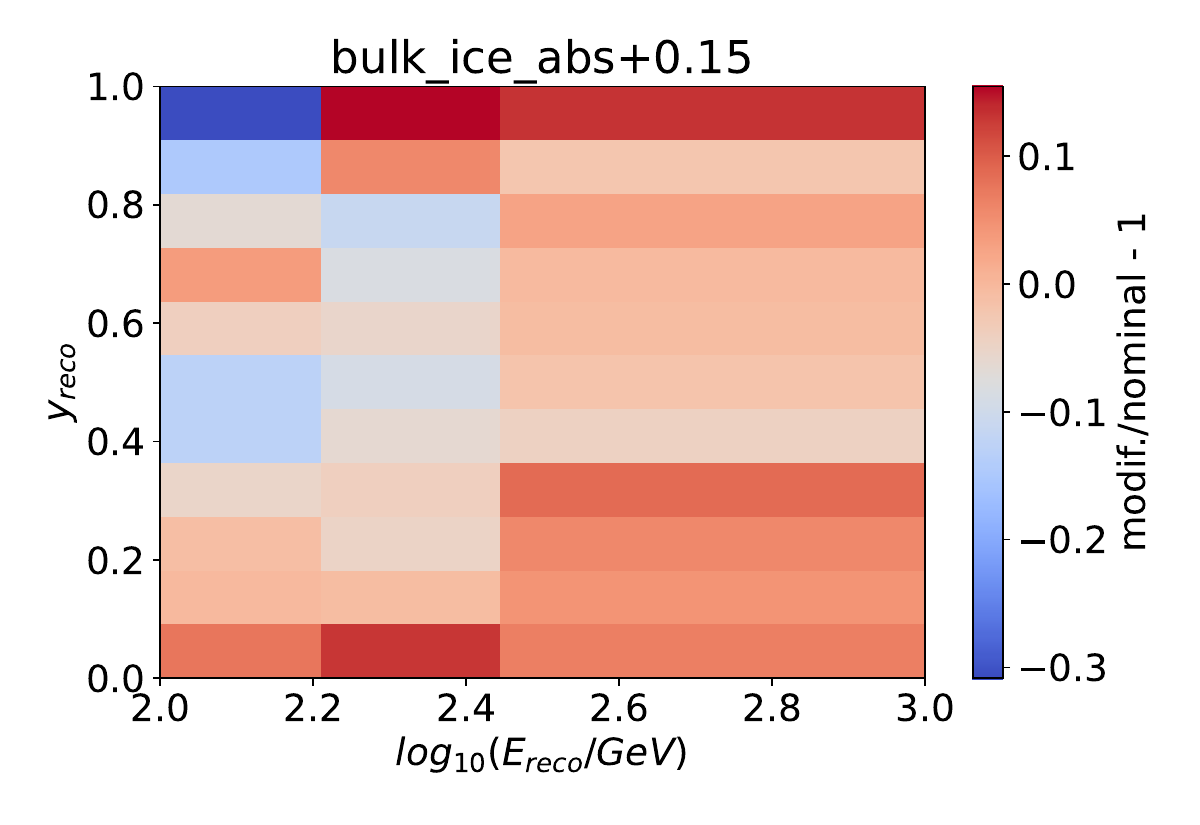}
        \includegraphics[width=0.31\linewidth]{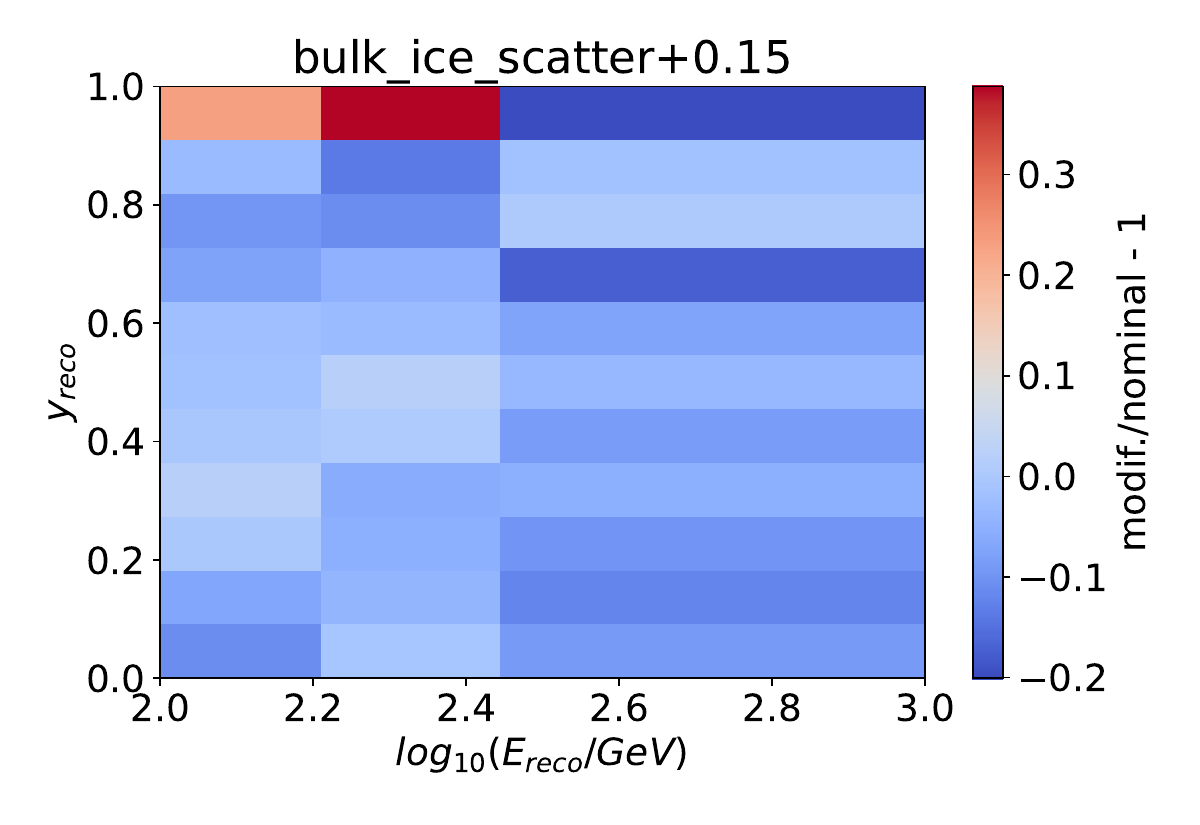}
        \includegraphics[width=0.31\linewidth]{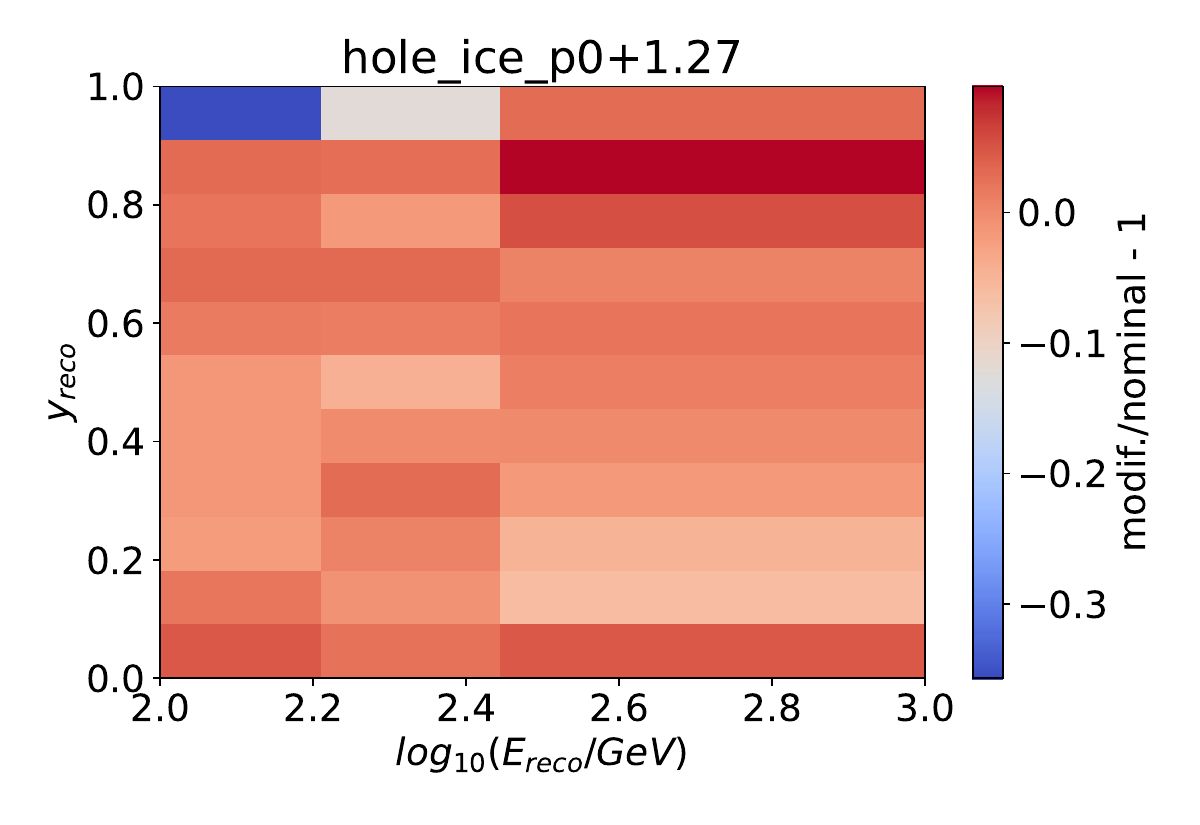}
        \includegraphics[width=0.31\linewidth]{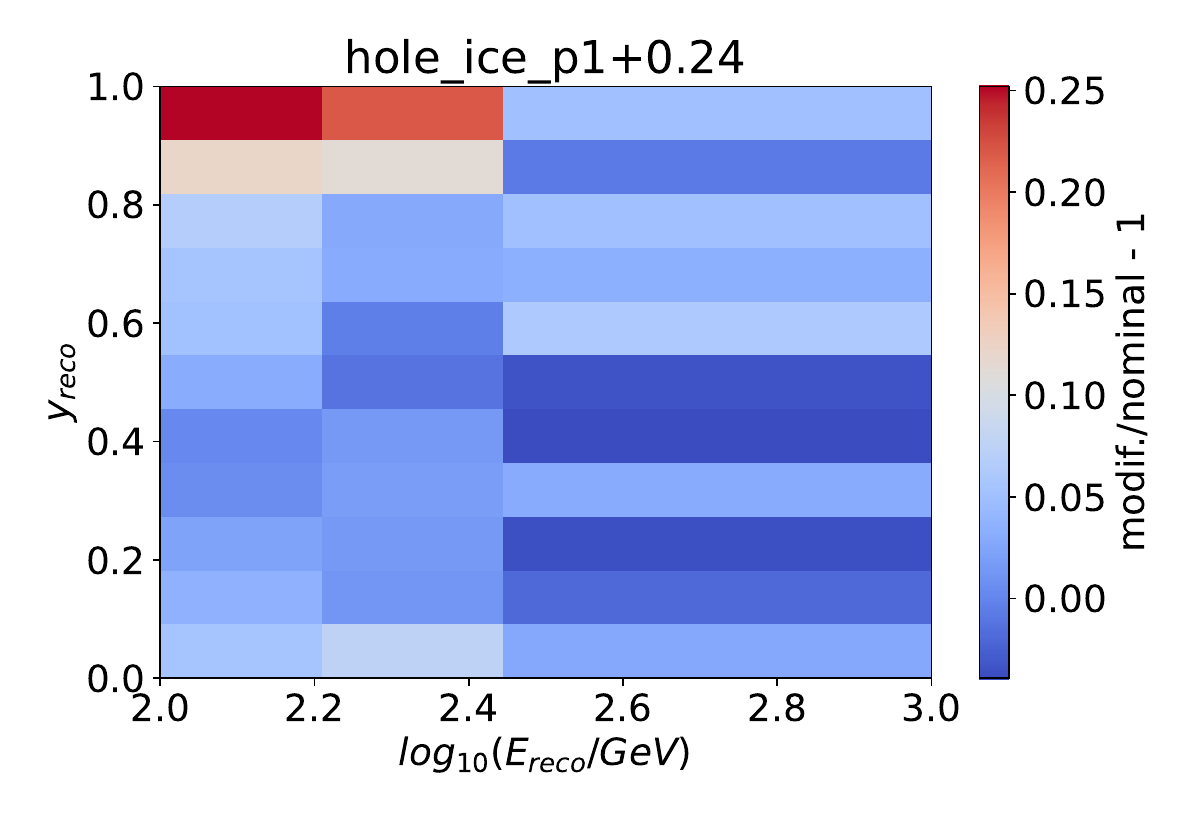}
    \caption{Fractional change in the number of events compared to the nominal expectation, introduced when each nuisance parameter was increased by $1\sigma$.}
    \label{fig::syst::syst_modif_to_templ}
\end{figure*}

In our parametrization, the {\sc genie}+HKKMS'15 model is best described by \meany$ = 0.43\pm0.02$ and $\log_{10}\lambda = 0.00^{+0.11}_{-0.08}$ in the energy range of interest. To illustrate the impact of assuming different values of \meany and $\lambda$, we calculate the expected change in number of events in the analysis binning of reconstructed energy and inelasticity. Figure~\ref{fig:example_dpdy} shows three example \dpdy distributions with different values of \meany and $\lambda$.
The comparison between the event counts corresponding to all three distributions and the expectation from {\sc genie}+HKKMS'15 is shown on Fig.~\ref{fig:app:maps_modif}. 
Figure~\ref{fig::syst::syst_modif_to_templ} shows changes in the number of events due to modifications in systematic parameter values, while the {\sc genie}+HKKMS'15 model is assumed for the inelasticity distribution.

\begin{figure*}
    \centering
    \includegraphics[width=0.7\linewidth]{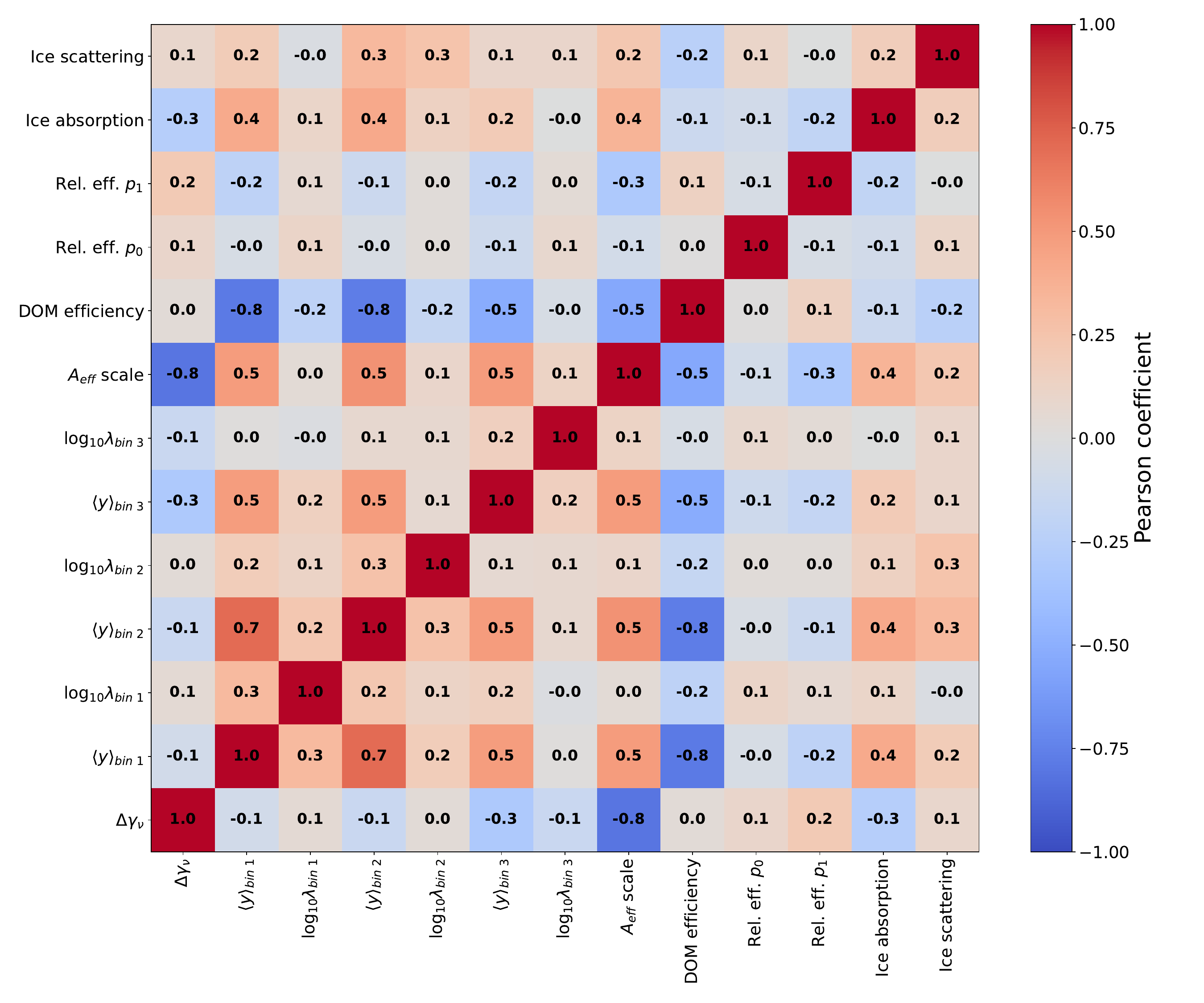}
    \caption{Parameter correlations based on ensemble test at the best-fit point.}
    \label{fig:app:correlations}
\end{figure*}

To quantify correlations between physics and nuisance parameters in the fit, we performed a pseudotrial test. We generated an ensemble of 400 pseudoexperiments assuming nominal values of fit parameters, but including Poisson fluctuations in each analysis bin. For each of the pseudoexperiments an inelasticity distribution fit was performed, and resulting posterior parameter distributions were used to calculate Pearson correlation coefficients for each pair of fit parameters.
Figure~\ref{fig:app:correlations} shows the Pearson correlation matrix based on all 400 trials in the test.

\begin{figure*}
    \centering
    \includegraphics[width=0.65\linewidth]{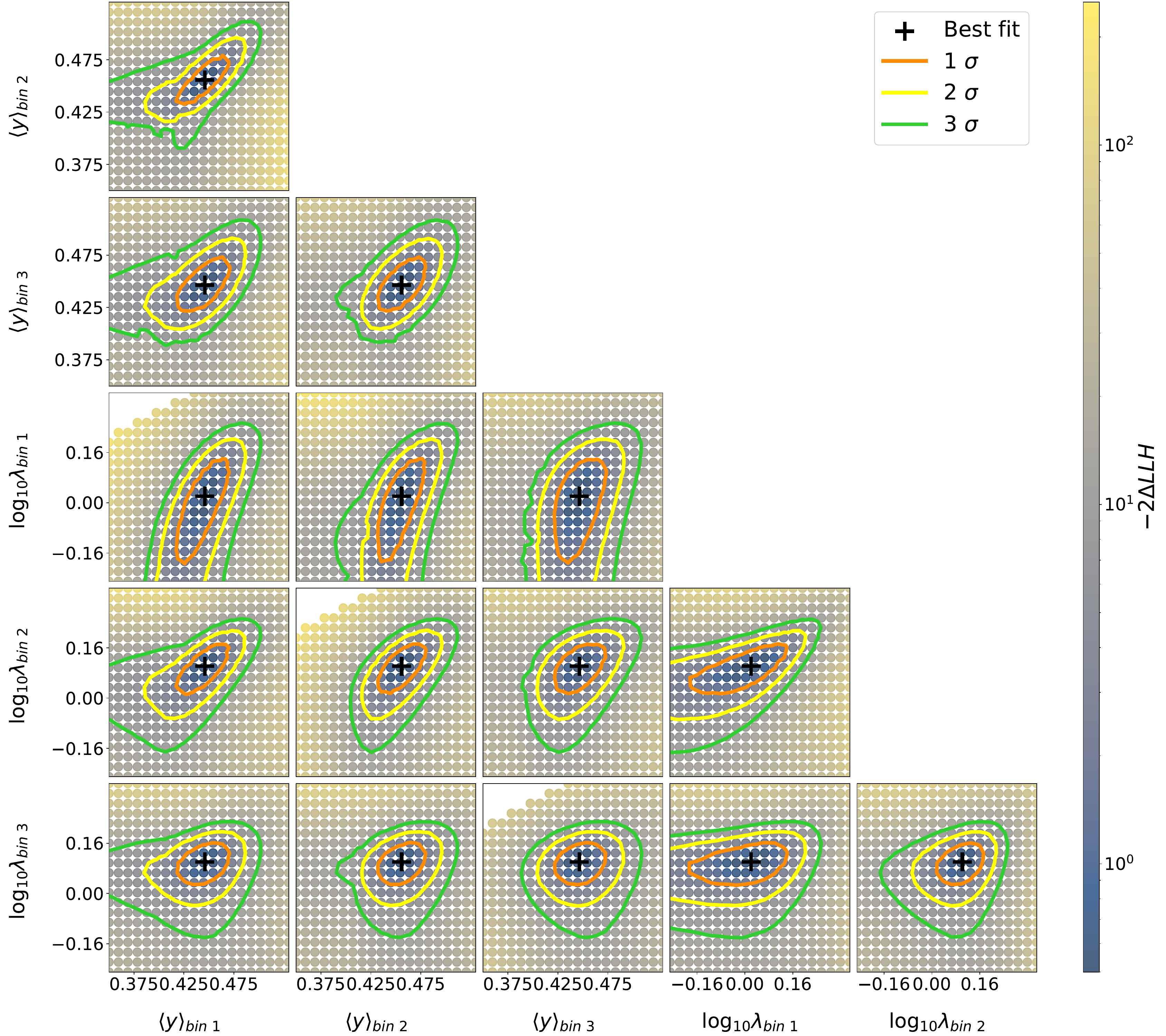}
    \caption{Two-dimensional likelihood scans for all unique combination of physics parameters. The black cross corresponds to the best-fit point and the contours show 1$\sigma$, 2$\sigma$ and 3$\sigma$ confidence intervals.}
    \label{fig:app:triangle_llh_scan}
\end{figure*}

After the fit to data were done, we also conducted a two-dimensional likelihood scan for all 15 unique pairs of physics parameters. Figure~\ref{fig:app:triangle_llh_scan} shows two-dimensional likelihood scans for each pair of parameters, featuring best-fit point and confidence level contours.

\clearpage
\bibliography{bib}

\end{document}